\documentclass[aps, prb, twocolumn, superscriptaddress]{revtex4-2}

\usepackage[utf8]{inputenc}
\usepackage[T1]{fontenc}						
\usepackage{lmodern}							
\usepackage{graphicx}   						
\usepackage{lineno}	[pagewise]					
\modulolinenumbers[5]							
\usepackage{xspace}
\usepackage{bm}								
\usepackage[normalem]{ulem}       
\usepackage{amsmath}
\usepackage[colorinlistoftodos]{todonotes} 
\usepackage[colorlinks, linkcolor = blue, citecolor = blue, filecolor = black, urlcolor = blue]{hyperref}
\usepackage{siunitx}
\DeclareSIUnit\Oersted{Oe}
\DeclareSIUnit\rlu{r.\,l.\,u.}
\DeclareSIUnit\barn{barn}
\usepackage[final]{changes}

\begin{document}

%%%%%%%%%%%%%%%%%%%%%%%%%%%%%%%%%
%Commands
\newcommand{\total}[2]{\ensuremath{\frac{\D #1}{\D #2}}}
\newcommand{\dell}[2]{\ensuremath{\frac{\partial #1}{\partial #2}}}
\newcommand{\mb}[1][B]{\ensuremath{\mu_{\textrm{#1}}}}
\newcommand{\tm}[1]{\textrm{#1}}
\newcommand{\te}{$\tilde{\epsilon}$}
\newcommand{\I}[1]{\textit{#1}}
\newcommand{\Vb}[1]{\ensuremath{\bm{#1}}}	%bold vector
\newcommand{\red}[1]{\textcolor{red}{#1}}
\newcommand{\green}[1]{\textcolor{olive}{#1}}
\newcommand{\blue}[1]{\textcolor{blue}{#1}}
\newcommand{\orange}[1]{\textcolor{orange}{#1}}
\newcommand{\magenta}[1]{\textcolor{magenta}{#1}}
\newcommand{\peak}[3]{(#1\,#2\,#3)}
\newcommand{\D}{\ensuremath{\mathrm{d}}}
\newcommand{\Rho}{\varrho}
\newcommand{\eg}{\textit{e.\,g.}\xspace}
\newcommand{\ie}{\textit{i.\,e.}\xspace}
\newcommand{\vs}{\textit{vs.}\xspace}
\newcommand{\kvec}[1]{\SI{#1}{\angstrom}\ensuremath{^{-1}}}
\newcommand{\etal}[1]{#1~\emph{et\,al.}\xspace}
\newcommand{\Sub}[2]{\ensuremath{#1_{\mathrm{#2}}}}
\newcommand{\fig}[1]{Fig.~#1}
\newcommand{\tab}[1]{Tab.~#1}
\newcommand{\tb}[1]{Tab.~#1}
\newcommand{\X}{\raisebox{2pt}{\ensuremath{\chi}}}
\newcommand{\cro}{\mbox{Ca$_2$RuO$_4$}\xspace}
\newcommand{\sro}{\mbox{Sr$_2$RuO$_4$}\xspace}
\newcommand{\bicro}{\mbox{Ca$_3$Ru$_2$O$_7$}\xspace}
\newcommand{\bisro}{\mbox{Sr$_3$Ru$_2$O$_7$}\xspace}
\newcommand{\new}[1]{\textcolor{blue}{#1}}
\newcommand{\rem}[1]{\textcolor{blue}{\sout{#1}}} 
%%%%%%%%%%%%%%%%%%%%%%%%%%%%%%%%%
%Sections

%-------------------------------------------
% TITLE & ABSTRACT

% Title 
\title{Critical magnetic fluctuations in the layered ruthenates Ca$_2$RuO$_4$ and Ca$_3$Ru$_2$O$_7$}

\author{H. Trepka}
\author{T. Keller}
\affiliation{Max-Planck-Institute for Solid State Research, Heisenbergstra{\ss}e 1, 70569 Stuttgart, Germany}
\affiliation{Max Planck Society Outstation at the Heinz Maier-Leibnitz Zentrum (MLZ), Lichtenbergstra{\ss}e 1, 85748 Garching, Germany}

\author{M. Krautloher}
\affiliation{Max-Planck-Institute for Solid State Research, Heisenbergstra{\ss}e 1, 70569 Stuttgart, Germany}

\author{J. Xu}
\affiliation{Helmholtz-Zentrum Berlin für Materialien und Energie, Hahn-Meitner Platz 1, D-14109 Berlin, Germany}
\author{K. Habicht}
\affiliation{Helmholtz-Zentrum Berlin für Materialien und Energie, Hahn-Meitner Platz 1, 14109 Berlin, Germany}
\affiliation{Institut für Physik und Astronomie, Universit{\"a}t Potsdam, Karl-Liebknecht-Stra{\ss}e 24-25, 14476 Potsdam, Germany}

\author{M. Böhm}
\affiliation{Institut Laue-Langevin, 71 Avenue des Martyrs, 38042 Grenoble Cedex 9, France}

\author{B. Keimer}
\email[]{B.Keimer@fkf.mpg.de}
\affiliation{Max-Planck-Institute for Solid State Research, Heisenbergstra{\ss}e 1, 70569 Stuttgart, Germany}
\author{M. Hepting}
\email[]{Hepting@fkf.mpg.de}
\affiliation{Max-Planck-Institute for Solid State Research, Heisenbergstra{\ss}e 1, 70569 Stuttgart, Germany}
\affiliation{Max Planck Society Outstation at the Heinz Maier-Leibnitz Zentrum (MLZ), Lichtenbergstra{\ss}e 1, 85748 Garching, Germany}

%Abstract
\begin{abstract}

Materials realizing the XY model in two dimensions (2D) are sparse. Here we use neutron triple-axis spectroscopy to investigate the critical static and dynamical magnetic fluctuations in the square-lattice antiferromagnets Ca$_{2}$RuO$_{4}$ and Ca$_{3}$Ru$_2$O$_{7}$. We probe the temperature-dependence of the antiferromagnetic (AFM) Bragg-intensity, the $Q$-width, the amplitude, and the energy-width of the magnetic diffuse scattering in vicinity to the N\'eel temperature $\Sub TN$ to determine the critical behavior of the magnetic order parameter $M$, correlation length $\xi$, susceptibility $\X$, and the characteristic energy $\Gamma$ with the corresponding critical exponents $\beta$, $\nu$, $\gamma$, and $z$, respectively. We find that the critical behaviors of the single-layer compound \cro follow universal scaling laws that are compatible with predictions of the 2D-XY model. The bilayer compound \bicro is only partly consistent with the 2D-XY theory and best described by the three-dimensional Ising \mbox{(3D-I)} model, which is likely a consequence of the intra-bilayer exchange interactions in combination with an orthorhombic single-ion anisotropy.
Hence, our results suggest that layered ruthenates are promising solid-state platforms for research on the 2D-XY model and the effects of 3D interactions and additional spin-space anisotropies on the magnetic fluctuations. 

\end{abstract}

\maketitle

%-------------------------------------------
% INTRODUCTION

\section{Introduction}

Critical fluctuations of the order parameter emerge in proximity to the transition temperature $\Sub Tc$ of second-order phase transitions.
These fluctuations are characterized by a correlation length $\xi$ and a
response time $\tau$, which diverge at $\Sub Tc$ \cite{Fisher.1967,
Stanley.1971, Hohenberg.1977, Collins.1989}. In the critical regime
close to $\Sub Tc$, fundamental physical properties of a material, such
as the magnetic susceptibility and the heat capacity, adopt critical
behavior and can be described by power-laws  $\propto |t|^\lambda$, with
critical exponents $\lambda$ and $t \equiv (T/T_c - 1)$
\cite{Collins.1989, Halperin.1969, Stanley.1971, Hohenberg.1977,
Pelissetto.2002}. Furthermore, the scaling behaviors in the spatial and time domains are related via $\Gamma \propto \kappa^z \propto t^{z\nu}$
\cite{Hohenberg.1977}, with the critical exponents $\nu$ of the inverse 
correlation length $\kappa = \xi^{-1}$ and $z$ of the characteristic energy $\Gamma \propto \tau^{-1}$. 
A hallmark of the corresponding scaling theory is the
concept of universality \cite{Griffiths.1970, Kadanoff.1993b, Collins.1989}, which stipulates that (for systems with short-range interactions) the critical exponents are independent of microscopic details, and depend exclusively on the dimensionality of space and the dimensionality of the order parameter.
In magnetic systems, the scaling behavior of magnetic critical fluctuations thus encodes the spatial dimensionality of the system and possible magnetic anisotropies \cite{Collins.1989}. Along these lines, in particular the 2D-XY model has attracted significant attention, since it was employed as the model system for the unconventional vortex-unbinding transition proposed by Berezinskii, Kosterlitz, and Thouless (BKT) \cite{Bereszinskii.1972, Kosterlitz.1973, Kosterlitz.1974}. The fingerprints of BKT-transitions were observed in superfluid $^4$He films \cite{Bishop.1978, Kosterlitz.2020} and proximity-coupled Josephson junction arrays \cite{Resnick.1981,Leemann.1986}. Yet, solid-state materials that realize the 2D-XY model are sparse \cite{L.P.Regnault.20, Hirakawa.1982, Bramwell.1995, Ronnow.2000,  Klyushina.2021}.

A key experimental technique for the investigation of critical magnetic
scattering is neutron triple-axis spectroscopy (TAS), which exploits the
proportionality between the magnetic neutron scattering cross section
and the dynamic scattering function $S(\Vb{q},\omega)$, containing $\kappa$
and $\Gamma$ \cite{Shirane.2002,Chatterji.2006,Squires.1996}. More
specifically, $\Gamma$ can be derived from TAS energy scans of the
critical magnetic scattering, while $\kappa$ corresponds to the
energy-integrated $Q$-width in momentum space. Along these lines, pioneering studies
investigated the critical magnetic fluctuations in classical magnetic systems,
such as the 3D ferromagnet (FM) EuO \cite{AlsNielsen.1976,
Dietrich.1976, Boni.1986} and the 3D antiferromagnet
(AFM) RbMnF$_3$ \cite{Tucciarone.1971, Coldea.1998}.  
Furthermore, TAS studies were carried out on systems with quasi-2D magnetic correlations, including the isotropic square-lattice AFMs Rb$_2$MnF$_4$ \cite{Christianson.2001}, Sr$_2$CuO$_2$Cl$_2$, and Sr$_2$Cu$_3$O$_4$Cl$_2$ \cite{Kim.2001}, as well as the AFM parent compounds of the cuprate superconductors \cite{Keimer.1992}, which exhibit 2D Heisenberg (2D-H) scaling properties above their N\'eel temperatures. 
More recently, critical
magnetic fluctuations were investigated in 5$d$-electron transition metal oxides (TMOs) using X-ray scattering. In single-layer Sr$_{2}$IrO$_{4}$, which exhibits a Mott-insulating AFM ground state with a $\Sub J{eff} = 1/2$ effective total angular momentum due to strong spin-orbit coupling (SOC) \cite{Kim.2008, Kim.2009}, 2D-H scaling with a small easy-plane anisotropy was reported \cite{Fujiyama.2012,Vale.2015}. 
On the other hand, in bilayer Sr$_{3}$Ir$_2$O$_{7}$ the scaling behavior close to the transition is consistent with the 3D Ising (3D-I) universality class, but significant deviations were found and attributed to disorder \cite{Vale.2019}.

\begin{figure*}%[htbp]
	\begin{center}
	\includegraphics[width=\textwidth]{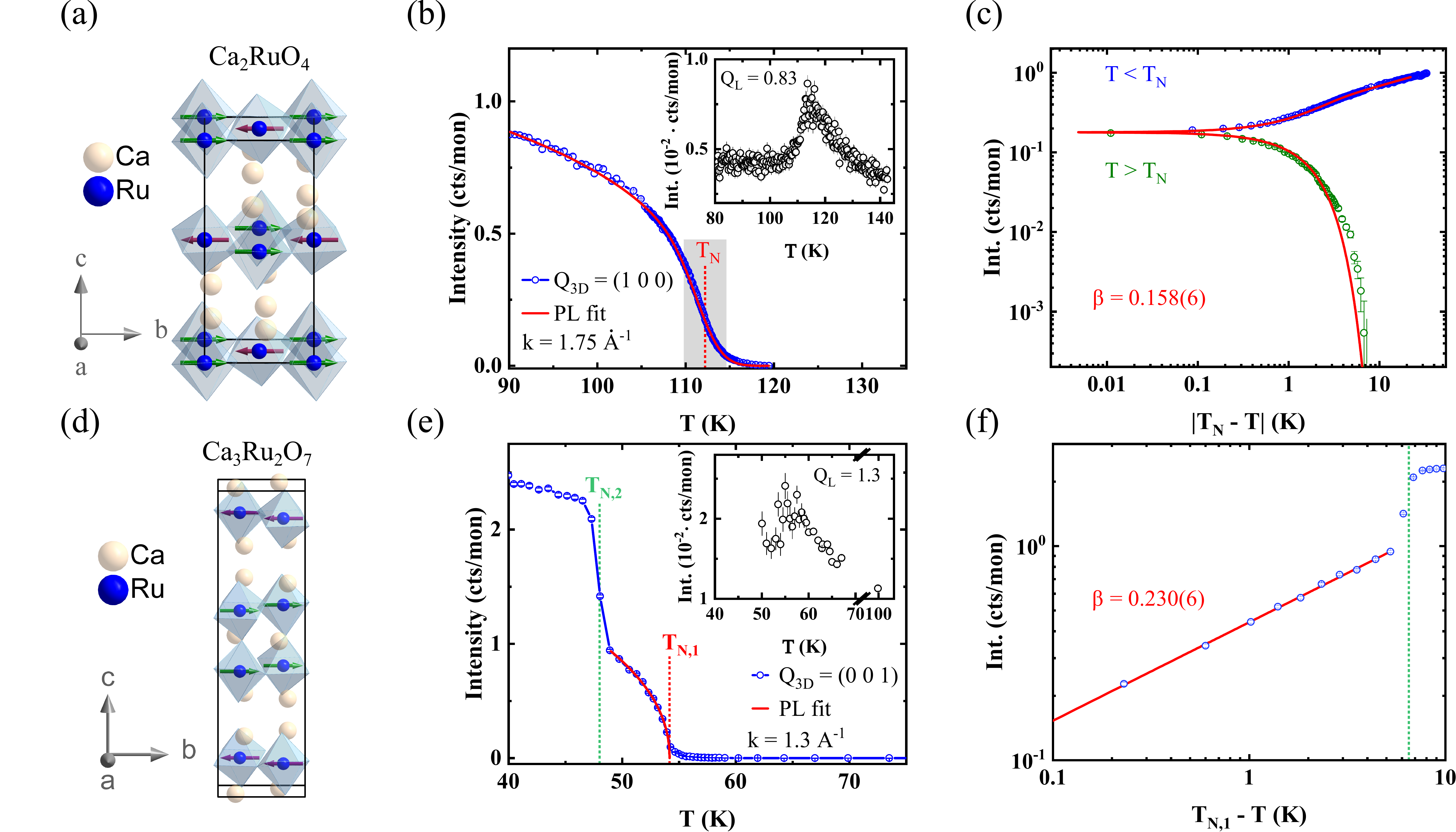}
		\caption{Crystal structure, critical scattering and magnetic order parameter in  \cro (a-c) and \bicro (d-f). (a) Schematic of the crystallographic unit cell of single-layer \cro (black lines). Oxygen ions are omitted for clarity. Green and purple spins indicate the G-type AFM order. (b) Intensity of the magnetic \peak100 peak measured as a function of temperature. The red solid line is a power-law (PL) fit $I \propto M^2 \propto |t|^{2\beta}$, with $\beta = 0.158(6)$ and $\Sub TN = \SI{112.20(1)}{\K}$, convoluted with a Gaussian-distribution of N\'eel temperatures \Sub TN with a FWHM of $\SI{4.84(1)}{\K}$ (grey shaded area). The inset shows the intensity measured at $Q = \peak10{0.83}$. The increase of intensity in vicinity to $\Sub TN$ indicates the presence of critical scattering from critical magnetic fluctuations. (c) Same data as in panel (b), but on a double-logarithmic scale. Note that the data and fits are non-linear due to the distribution of \Sub TN. (d) Schematic of the unit cell of bilayer \bicro (black lines) with the A-Type AFM order indicated. (e) Intensity of the magnetic \peak001 peak. The green and red dashed lines indicate the two AFM transitions. Below $\Sub T{N,2} = \SI{48}{\K}$, the magnetic moments reorient from $m \parallel a$ to $m \parallel b$. The red solid line is a PL fit, which yields $\Sub T{N,1} = \SI{54.16(2)}{\K}$ and $\beta = 0.230(6)$. The inset shows the intensity measured at $Q = \peak00{1.3}$, with the increase of intensity in vicinity to $\Sub TN \sim \SI{54}{\K}$ indicating the presence of critical scattering. (f) Same data for $T < \Sub T{N,1}$ as in panel (e), but on a double-logarithmic scale.}
		\label{fig:results:Tscan}
	\end{center}
\end{figure*}

In 4$d$-electron TMOs, such as single- and bilayer ruthenates, critical fluctuations have remained unexplored to date. Notably, ruthenates show a plethora of electronic ground states \cite{Cao.2013, Markovic.2020, Grigera.2001, Horio.2021, Sidis.1999} such as unconventional superconductivity in \sro \cite{Maeno.1994} and excitonic AFM order in the Mott insulator \cro \cite{Khaliullin.2013, Akbari.2014}, arising from a delicate competition between the energy scales of SOC, crystal field
splitting, Hund's coupling, and inter-site exchange interactions. In the latter compound, spins are arranged in an AFM fashion within square-lattice RuO$_2$ planes and stacked along the $c$-axis in a G-type
pattern [\fig{\ref{fig:results:Tscan}}a] with a N\'eel temperature $\Sub TN \sim \SI{110}{\K}$ \cite{Braden.1998, Alexander.1999, Nakatsuji.2001}. The excitonic character is believed to result from excitonic
transitions between non-magnetic singlet ($\Sub J{eff}$=0) and
magnetic triplet states ($\Sub J{eff}$=1)  \cite{Khaliullin.2013,
Akbari.2014}. The nature of the excitonic magnetism was
recently corroborated by resonant inelastic x-ray scattering (RIXS)
\cite{Gretarsson.2019}, Raman scattering \cite{Souliou.2017}, as well as inelastic neutron scattering (INS), detecting a soft amplitude mode ('Higgs-mode') in the spin-wave spectrum
\cite{Jain.2017}.

The unquenched orbital angular momentum of the Ru magnetic moments in \cro further results in a highly unusual spectrum of transverse magnons in the AFM state \cite{Jain.2017}. The low-energy magnetic Hamiltonian derived from an analysis of this spectrum is dominated by an XY-type single-ion anisotropy, which is much larger than the nearest-neighbor exchange interaction and an Ising-type single-ion anisotropy resulting from an orthorhombic distortion of the crystal structure. At the same time, the INS experiments did not reveal any dispersion of the magnons perpendicular to the RuO$_2$ layers, which implies that the interlayer interactions are much weaker than the interactions within the layers. The evidence for an approximate 2D-XY symmetry of the magnetic Hamiltonian derived from the analysis of the magnon dispersions has motivated the present study.

In contrast to the Mott insulator \cro, the bilayer compound \bicro   is metallic in the paramagnetic state and maintains considerable electrical conductivity below the N\'eel temperature $\Sub T{N,1} \sim 56$\,K \cite{McCall.2003}. The magnetic structure is A-type AFM (\ie FM bilayers with alternating orientation along the $c$-axis) [\fig{\ref{fig:results:Tscan}}d] \cite{McCall.2003,Yoshida.2005}. A second magnetic transition associated with a reorientation of the spins from the $a$- to the $b$-axis in the RuO$_2$ planes \cite{Bohnenbuck.2008} and a greater reduction of the electrical conductivity occurs at $\Sub T{N,2} \sim 48$\,K \cite{McCall.2003}. As the crystal structure of \bicro comprises two closely spaced RuO$_2$ layers within a unit cell [\fig{\ref{fig:results:Tscan}}d], substantial interlayer interactions within a bilayer unit are expected and were indeed identified in INS studies of the magnon dispersions \cite{Ke.2011b,Bertinshaw.2021}. As exchange interactions between bilayer units are weak, the dimensionality of the exchange-bond network is intermediate between 2D and 3D. The INS data also revealed an anisotropy gap, but were insufficient for a determination of the nature of the dominant anisotropy (Ising versus XY).

In this work, we use TAS to examine the critical scattering in single-layer \cro and bilayer \bicro in vicinity and above $\Sub TN$. We extract the 
critical static and dynamical exponents to determine the spin dimensionalities and anisotropies, which we compare to the model Hamiltonians employed in previous INS and RIXS studies below $\Sub TN$. For \cro, we derive the critical exponent $\beta$ of the order parameter from the temperature-dependence of the AFM \peak100 Bragg intensity. The static critical exponents $\nu$ and $\gamma$ are extracted from the $Q$-width and amplitude, respectively, of the magnetic diffuse scattering around \peak10{0.83} above $\Sub TN$. We find that 
the temperature-dependence of the order parameter, $Q$-width, and amplitude are well-captured by a 2D-XY model, as expected based on the spin Hamiltonian extracted from the magnon dispersions \cite{Jain.2017}. The dynamic critical exponent $z$ is derived from the broadening in energy of the diffuse scattering at \peak100 for $T > \Sub TN$ and is also in reasonable agreement with the 2D-XY model. 
For \bicro, we derive $\beta$ from the temperature-dependence of the AFM \peak001 Bragg intensity below \Sub TN, while $\nu$, $\gamma$ and $z$ are extracted from the $Q$-width, amplitude, and energy-width, respectively, of the diffuse scattering around \peak001 above \Sub TN. 
From a combined consideration of all extracted exponents, we conclude that the critical behavior of \bicro is only partly consistent with the 2D-XY model, and is best described by the 3D-I model. We discuss these observations in the context of prior experiments on single-layer and bilayer iridates \cite{Fujiyama.2012, Vale.2015, Vale.2019}. 

%-------------------------------------------
% METHODS

\section{Methods}
High-quality single crystals of \cro and \bicro were grown by the optical floating zone method \cite{Nakatsuji.2001}, as described in Ref.~\cite{Jain.2017}. \cro exhibits the orthorhombic space group {\it Pbca} at \SI{11}{\K} and the lattice parameters $a=\SI{5.39}{\angstrom}$, $b=\SI{5.63}{\angstrom}$, and $c=\SI{11.75}{\angstrom}$ \cite{Braden.1998}. \bicro crystallizes in an orthorhombic space group {\it Bb2$_1$m} at \SI{50}{\K} with the lattice parameters $a=\SI{5.36}{\angstrom}$, $b = \SI{5.53}{\angstrom}$, and $c = \SI{19.54}{\angstrom}$  \cite{Yoshida.2005}. Single crystals that included orthorhombic ($a$,$b$)-twins were co-aligned on Si-plates with thicknesses of \SI{0.5}{\mm} and arranged in Al-sample holders. In case of \cro, approximately 100 single crystals were co-aligned, yielding a total mass of $\SI{1.5}{\g}$. In case of \bicro, approximately 30 single crystals were used with total mass of  $\SI{0.8}{\g}$. The mosaicity of both sample arrays was 2-3\,°. Due to the ($a$,$b$)-twinning, the scattering planes were \peak H0L/\peak 0KL. All values of $Q$ are given in reciprocal lattice units.

The two-axis mode experiments on \cro were carried out at the thermal neutron spectrometer \mbox{TRISP} \cite{Keller.2002b, Keller.2015} at the \mbox{FRM II} neutron source at the Heinz Maier-Leibnitz Zentrum \mbox{(MLZ)}, Garching. The instrument was operated with clockwise scattering sense at the monochromator and sample ($SM = -1$, $SS = -1$) at $k_i = \kvec{1.75}$. \added{Additionally, first neutron spin-echo (NSE) measurements were conducted at $Q = \peak100$ with $k_i = 2.66\,\AA^{-1}$ (TAS energy resolution/vanadium width $\approx 1$\,meV) and the instrumental configuration $SM = -1$, $SS = -1$, and $SA = -1$.} The dynamic properties were measured at the cold neutron TAS \mbox{FLEXX} \cite{Le.2013} at the \mbox{BER II} neutron source at the Helmholtz-Zentrum Berlin \mbox{(HZB)}, Berlin. An instrumental configuration $SM = -1$, $SS = 1$, $SA = -1$, open collimation, and a neutron wave vector $k_f = \kvec{1.3}$ were used. The energy resolution was $\approx \SI{0.15}{\milli\electronvolt}$. A Be-filter in addition to a velocity selector in the primary spectrometer was used to suppress higher monochromator orders. A small offset in the thermometry between \mbox{TRISP} and \mbox{FLEXX} was corrected by comparing the \peak100 peak intensities.

The experiments on \bicro \cite{TREPKAHeiko.2020} were carried out at \mbox{ThALES} \cite{Boehm.2007}, which is a cold neutron TAS at the Institut Laue-Langevin \mbox{(ILL)}, Grenoble. The instrument was operated (i) in two-axis mode with $SM = 1$, $SS = -1$ and $k_i = \kvec{1.3}$ and (ii) in three-axis mode with $SM = 1$, $SS = -1$, $SA = 1$ and $k_f = \kvec{1.3}$ with double focusing monochromator and analyzer (energy resolution $\approx \SI{0.08}{\milli\electronvolt}$). A Be-filter was used to suppress higher monochromator orders. 

The cross section of magnetic neutron scattering \cite{Shirane.2002, Boothroyd.2020} is proportional to the dynamic scattering function $S(\Vb q,\omega)$, with $\Vb Q = \Vb G_m + \Vb q = \Vb k_i - \Vb k_f$ and $\omega = \hbar(k_i^2 - k_f^2)/(2m)$. 
Here, $\Vb G_m$ is a magnetic reciprocal  lattice vector, $\Vb q$ the relative momentum transfer, and $\Vb k_{i,f}$ the incident and final neutron wave vectors. $S(\Vb q,\omega)$ is related to the imaginary part of the generalized magnetic susceptibility via 
\begin{equation}
S(\Vb q, \omega) = \frac{\X''(\Vb q, \omega)}{1-\exp(-\hbar\omega/k_\tm BT)}.
\end{equation}
The real and imaginary parts of the generalized susceptibility $\X(\Vb q, \omega)$ are Kramers–Kronig related. A general form of $\X''(\Vb q, \omega)$ is given by $ \X''(\Vb q, \omega)= \X'(\Vb q)F(\omega)\omega$, where $\X'(\Vb q)$ is the real part of the static susceptibility and $F(\omega)$ the spectral weight function, which is an even function of $\omega$ and satisfies the normalization condition $\int_{-\infty}^\infty F(\omega) \D\omega = 1$. Above the ordering temperature, spin fluctuations at small $\Vb q$ are strongly damped and the spectral-weight function takes on a Lorentzian shape \cite{Shirane.2002}:   
\begin{equation}
F(\omega) = \frac{1}{\pi}\frac{\Gamma}{\Gamma^2 + \omega^2}.\\
\label{equ:methods:F_omega}
\end{equation}
To extract the inverse of the magnetic correlation length $\kappa$ from $Q$-scans, we use the following Lorentzian form for the static susceptibility \cite{Collins.1989}
\begin{equation}
    \X'(\Vb q)= \frac{\X'(\Vb 0)}{1 + q^2/\kappa^2}, 
\label{equ:methods:Ornstein}
\end{equation}
where $\X'(\Vb 0) \equiv \X_0$ corresponds to the staggered magnetic susceptibility. 
The Kramers-Kronig relation connects $\X'(\Vb q)$ and $S(\Vb q,\omega)$ via 
\begin{eqnarray}
\Sub kBT \X'(\Vb q) & = & \int_{-\infty}^{\infty}\frac{1-\exp(-\hbar\omega/k_\tm BT)}{\hbar\omega/\Sub kBT}S(\Vb q, \omega)~\D(\hbar\omega) \nonumber \\
& \simeq & \int_{-\infty}^{\infty} S(\Vb q, \omega)~\D(\hbar\omega) = S(\Vb q), 
\end{eqnarray}
where $\hbar\omega \ll \Sub kBT$ was assumed \cite{Shirane.2002,Boothroyd.2020}. Hence, $S(0) \propto \X_0T$ follows for the static case at $\Vb q = 0$. To determine S(\Vb q), in principle, it would be required to measure the entire S(\Vb q,$\omega$) function and perform a numerical $\omega$-integration, which can be avoided in 2D systems by using an energy integrating TAS configuration, as introduced by \etal{Birgeneau} \cite{Birgeneau.1971}. In this configuration the TAS analyzer is removed (two-axis mode) and $k_f$ is aligned perpendicular to the 2D-layers, corresponding to the \textit{ab}-plane in \cro. The magnitude of $k_f$ varies with $\omega$, but the relevant components of \Vb q in the 2D planes are constant and independent of $\omega$. In consequence, the detector signal corresponds to an energy integration with lower integration limit (energy-gain scattering) given by the thermal energy of the fluctuations, and upper limit (energy-loss scattering) given by the energy $E_i$ of the incident neutrons: 
\begin{equation}
S(\Vb q) = \int_{-\infty}^{\infty} S(\Vb q,\omega)~\D(\hbar\omega) \approx \int _{-k_BT}^{E_i} S(\Vb q,\omega)~\D(\hbar\omega). 
\label{equ:ScatTheo:approxBirg}
\end{equation}

In the case of \cro, we achieved this energy-integrating configuration with \mbox{$k_f \parallel c$} by choosing \mbox{$Q = \peak10{0.83}$} for $k_i = \kvec{1.75}$. For \bicro, due to the 3D character of the AFM order \cite{Bertinshaw.2021}, the ideal energy-integration configuration with \mbox{$k_f \parallel c$} can in principle not be obtained, as \Sub QL cannot be chosen arbitrarily. However, as discussed in App.~\ref{app:energyIntegrationBicro}, the effect of this imperfect energy-integration on the measured linewidth $\kappa$ in the two-axis configuration (without analyzer) is insignificant for our determination of the critical exponent $\nu$. Thus, the two-axis data of \bicro were not corrected for the integration effect.

%-------------------------------------------
% RESULTS

\section{Results}

\subsection{Static critical properties of \cro}

In \cro, the magnetic moments point along the $b$-axis of the orthorhombic unit cell
\cite{Braden.1998} [\fig{\ref{fig:results:Tscan}a}] with a possible  small canting in the $c$-direction ($m_c \approx 0.1m_b$) \cite{Porter.2018}. The magnetic susceptibility in the paramagnetic state indicates quasi-2D spin fluctuations \cite{Nakatsuji.1997}, which was recently also found from the magnon dispersion in the ordered phase \cite{Jain.2017}, where the 
following parameters were derived: $J = \SI{5.8}{\milli\electronvolt}$, 
$\Sub J{XY} = \SI{0.87}{\milli\electronvolt}$ for the Heisenberg and 
XY-type exchange couplings; and $E = \SI{25}{\milli\electronvolt}$, 
$\epsilon = \SI{4}{\milli\electronvolt}$ for the single-ion terms of the tetragonal and orthorhombic symmetries, respectively. An interlayer coupling $J'$ was not required to describe the magnon dispersion \cite{Jain.2017}, which is in line with studies on 1\% Ti-doped \cro, where a very small $J' = 0.03$\,meV 
was reported \cite{Kunkemoller.2015}. Hence, the strong tetragonal term $E$ 
and the small $J'$ signal that \cro can be regarded as a quasi-2D-XY AFM.

In the following TAS measurements on \cro, we use the energy-integrating two-axis mode (see Methods and Ref.~\cite{Birgeneau.1971}), which can be applied due to the 2D-character of the magnetism, with critical fluctuations that are expected to be independent of \Sub QL. We perform \Sub QH-scans around $Q = \peak10{0.83}$, which lies on the rod of the 2D magnetic scattering intensity. This corresponds to an energy-integrating configuration with alignment of $k_f 
\parallel c$ at $Q = \peak10{0.83}$ for $k_i = \kvec{1.75}$. Moreover, the advantage of a momentum $Q$ that is slightly off from a magnetic Bragg peak position is that the signatures of critical scattering can be particularly pronounced \cite{Dietrich.1969}. Accordingly, we observe an enhancement of the scattered intensity at $Q = \peak10{0.83}$ [inset in \fig{\ref{fig:results:Tscan}b}] for 
temperatures in vicinity to the anticipated \Sub TN of approximately 110\,K \cite{Nakatsuji.1997,Braden.1998}. More specifically, we observe that the critical scattering intensity peaks at a temperature slightly higher than 110\,K. This behavior is likely related to the fact that \Sub TN of our \cro sample is not sharply defined, but a  distribution of N\'{e}el temperatures is present
[see gray shaded area in \fig{\ref{fig:results:Tscan}}b], in spite of the confirmed excellent crystalline quality (see Methods and Ref.~\cite{Jain.2017}). This variance of \Sub TN likely results from microstrains within the crystal, which emerge below the concomitant structural and metal-to-insulator transition at 360\,K \cite{Alexander.1999} and could be reminiscent of the (pseudo)spin-lattice coupling in Sr$_2$IrO$_4$ \cite{Lupascu.2014,Porras2019}. In the following analysis, we take this distribution of \Sub TN into account, which allows us to extract the critical properties of \cro similarly to the case of a sharply defined \Sub TN. 

For an ideal second order phase transition, the order parameter (staggered magnetization $M$) is $I \propto M^2 \propto |t|^{2\beta}$ and vanishes above \Sub TN. Thus, the critical exponent $\beta$ can be determined from the measured nominal magnetic \peak100 peak intensity $I_{100}$ in \fig{\ref{fig:results:Tscan}}b. In general, $\beta$ and the other critical exponents are extracted from the slopes of linear fits in double-logarithmic plots (see \bicro below). However, due to the present variance of \Sub TN, the 
data in \fig{\ref{fig:results:Tscan}}b cannot be described directly with the power law (PL)  
scaling function. Note that especially the intensity around \SI{110}{\K} does not show the 
expected sharp drop but is smeared out. This rounding of the intensity evolution cannot be attributed to critical scattering above 
\Sub TN, since the data in the inset of \fig{\ref{fig:results:Tscan}}b indicate that the critical contribution is two orders of magnitude smaller. 
Thus, we fit the 
\peak100 data in the range 90 - 120\,K ($-0.2 < t < 0.1$) with a convolution of the above mentioned PL and a Gaussian distribution of $\Sub TN$ with full width at half maximum (FWHM) $\Delta\Sub TN$  [\fig{\ref{fig:results:Tscan}}b]. The resulting fit parameters are $\Sub TN = 
\SI{112.20(1)}{\K}$, $\Delta\Sub TN = \SI{4.84(1)}{\K}$, and $\beta = 
0.158(6)$. The $\beta$-value lies in between the limits of the 2D Ising 
(2D-I) ($\beta = 0.125$ \cite{Collins.1989}) and 2D-XY model ($\beta = 0.23$ 
\cite{Bramwell.1993,Bramwell.1993b}), as suggested for a XY system with fourfold crystal 
field anisotropy (XY$h_4$) \cite{Taroni.2008}. Figure~\ref{fig:results:Tscan}c shows $I_{100}$ and the fit curve on a double-logarithmic scale, illustrating that the PL fit provides an adequate description of the data below \Sub TN, and also for a range of temperatures above \Sub TN. Note that the strong deviation from a simple 
PL (straight line in double logarithmic plot) is due to the Gaussian 
distribution of \Sub TN. 

\begin{figure}[htbp]
     \begin{center}
\includegraphics[width=\columnwidth]{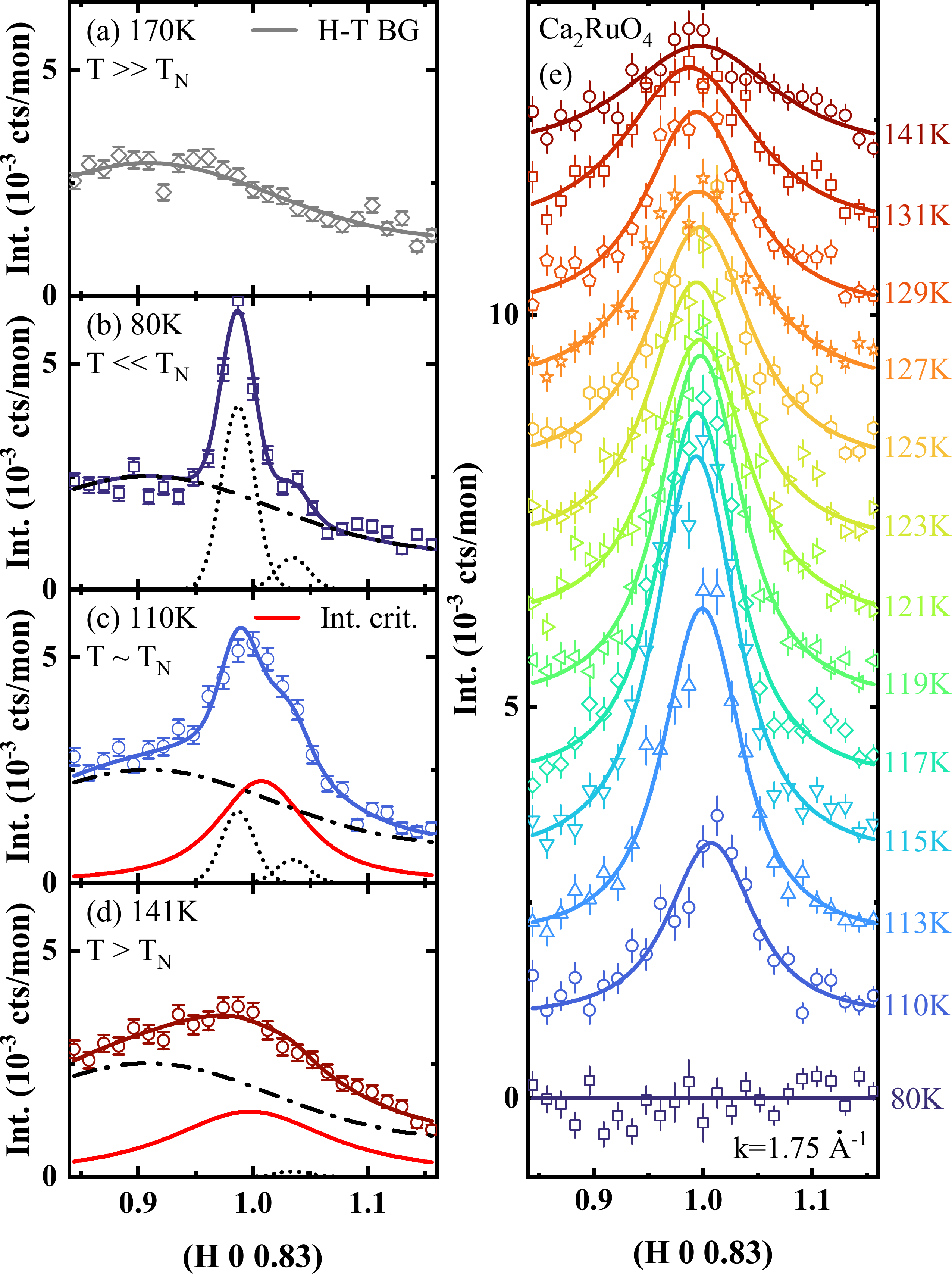}
         \caption{Selected energy-integrated \Sub QH-scans around \peak{H}0{0.83} for \cro before (a-d) and after (e) background (BG) subtraction. The 
data contain a $T$-independent constant BG, which was determined at $T = 
170\,K$ and labelled as 'high-temperature' (H-T) BG (a). Panels (b-d) 
show the H-T BG as a dashed-dotted line. At low-$T$, two small $T$-dependent 
resolution limited Gaussian-peaks (black dotted lines) appear (b-d), 
which are attributed to the magnetic \peak101 peak of the main domain and the \peak011 peak of the twin domain, 
respectively. The critical scattering component (Voigt-profile) is shown 
as a red line. (e) Selected \Sub QH-scans after BG subtraction 
with corresponding fit functions (solid lines). For clarity the data are 
plotted with a constant offset.}
                 \label{fig:results:QHscans_CRO214_examples}
     \end{center}
\end{figure}

As a next step we perform \Sub QH-scans around \peak{1}0{0.83} to 
determine the inverse correlation length $\kappa(T)$ 
[\fig{\ref{fig:results:QHscans_CRO214_examples}}]. Prior to fitting of 
the scans with Voigt-profiles, we thoroughly determine the background 
(BG) contributions. Representative scans are shown in
\fig{\ref{fig:results:QHscans_CRO214_examples}}a-d. We identify several 
components of the BG: (i) A temperature-independent component, which is 
determined at 170\,K [\fig{\ref{fig:results:QHscans_CRO214_examples}}a], 
\textit{i.e.} well above \Sub TN. The obtained fit (H-T BG) is employed 
as BG in the analysis of the data measured at all other temperatures 
(see dashed-dotted lines in 
\fig{\ref{fig:results:QHscans_CRO214_examples}}a-d). (ii) The scan at 
the lowest measured temperature $T = 80$\,K 
[\fig{\ref{fig:results:QHscans_CRO214_examples}}b] shows two 
incommensurate peaks besides the H-T BG. By comparing the 
temperature-dependent intensity of these resolution limited peaks with 
the intensities of the magnetic \peak100 
[\fig{\ref{fig:results:Tscan}}c] and \peak101 peaks \cite{Braden.1998}, 
they can be assigned to the \peak101 peak of the main domain, and the 
\peak011 peak of the twin domain \cite{Porter.2018}. The \peak101 peak is likely 
associated with a 'B-centered' phase with a different propagation vector 
and transition temperature $\Sub T{N,101} \approx \SI{150}{\K}$ 
\cite{Braden.1998}. Finally, we subtract the H-T BG and the 
aforementioned two peaks with proper $T$-scaling from the $Q$-scans and 
obtain the corrected data shown in 
\fig{\ref{fig:results:QHscans_CRO214_examples}}e. These data are well 
described by Voigt-profiles, which correspond to the convolution of the 
intrinsic Lorentzian with half-width-half-maximum (HWHM) $\kappa$ [Eqn.~\eqref{equ:methods:Ornstein}] and 
the Gaussian instrumental resolution. 
The constant width of the instrumental Gaussian (FWHM $\approx 0.034$\,r.l.u.) was extracted from the 80\,K scan in agreement with simulations carried out with the RESLIB \cite{Zheludev.2009} and TAKIN \cite{Weber.2016} softwares, respectively.

\begin{figure}[tb]
	\begin{center}
	\includegraphics[width=0.85\columnwidth]{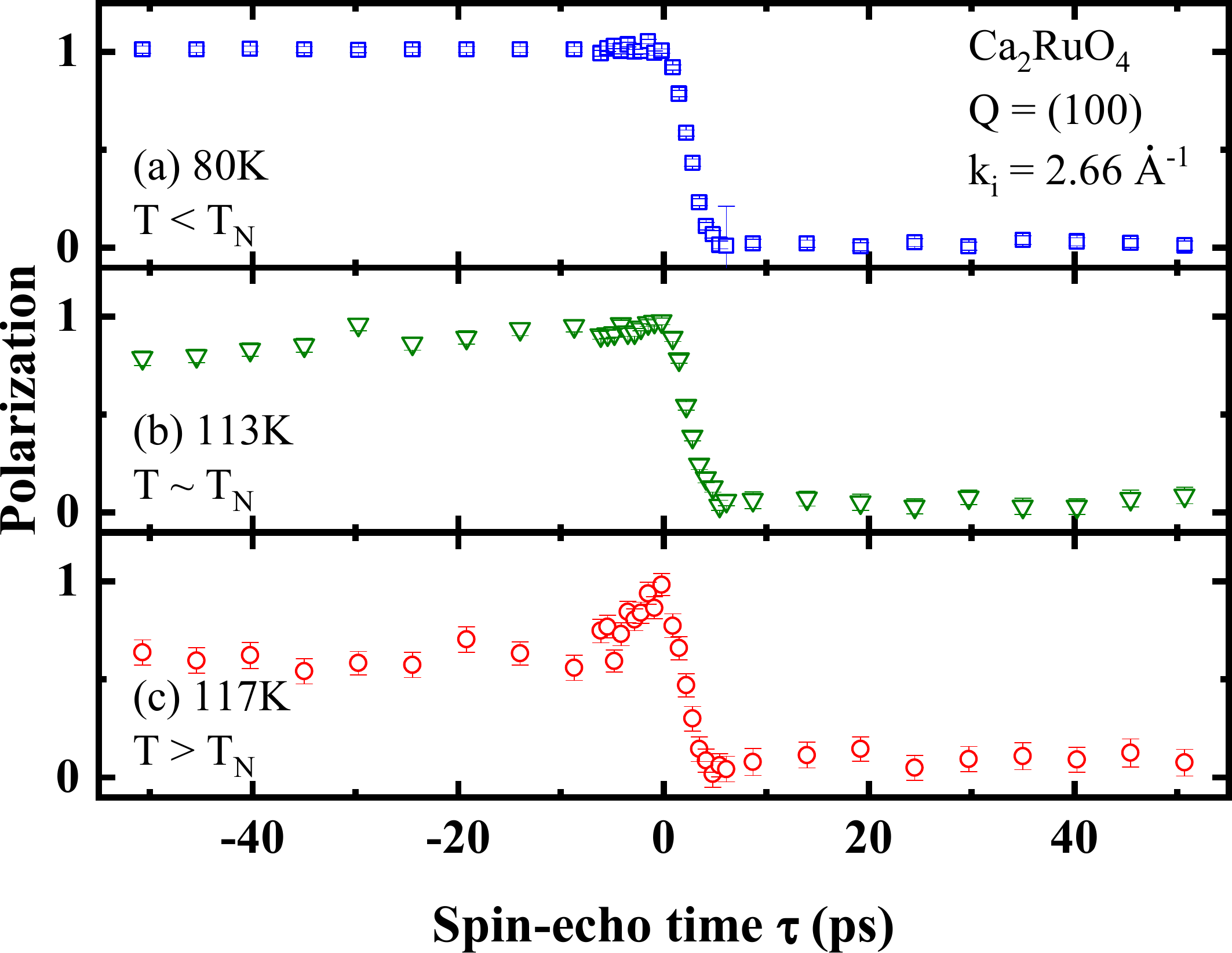}
		\caption{Neutron spin-echo (NSE) polarization \vs spin-echo time $\tau$ at $Q = \peak100$ for selected temperatures. According to Ref.~\cite{Tseng.2016}, we define $\tau < 0$ for measurements in spin-echo configuration, \ie with antiparallel magnetic fields $\Vb B_0$ in the spin-echo arms, and $\tau > 0$ for measurements in Larmor-diffraction mode, \ie  with parallel magnetic fields in the spin-echo arms. The lack of a significant polarization at $\tau > 0$ and the absence of oscillations of the polarization at small $\tau$ \cite{Tseng.2016} imply that the contribution of transverse fluctuations along the $c$-axis to the signal is marginal.}
		\label{fig:supp:CRO214_NSE}
	\end{center}
\end{figure}

Prior to the discussion of the inverse correlation length $\kappa$ extracted from  \fig{\ref{fig:results:QHscans_CRO214_examples}}e, we address the possible presence of concomitant longitudinal and transverse fluctuations. \added{In general, critical longitudinal (parallel to the static ordering vector) and non-critical transverse fluctuations are expected to be both visible and not separated for all measurement configurations used in this work. However, neutron spin-echo (NSE) spectroscopy \cite{Mezei.1980, Keller.2021, Mezei.1982, Mezei.1984} is capable to separate the two components \cite{Tseng.2016}. To this end, we carried out high-resolution NSE measurements on \cro at TRISP (see Methods), at $Q = \peak100$ with $k_i = 2.66\,\AA^{-1}$. Figure~\ref{fig:supp:CRO214_NSE} shows the resulting spin-echo polarization vs. the spin-echo time $\tau \equiv (m^2 \omega_L L)/(\hbar^2k_i^3)$ for selected temperatures, with the Larmor-frequency $\omega_L = \gamma_n \Vb B_0$ ($\gamma_n = 2.916$\,kHz/Oe) corresponding to the static magnetic fields $\Vb B_0$ in the spin-echo arms with length $L$. A high-temperature BG (170\,K) was subtracted from the data. According to Ref.~\cite{Tseng.2016}, possible transverse fluctuations in \cro (along the $c$-axis) would lead to a polarized signal for $\tau > 0$, \ie for parallel magnetic fields $\Vb B_0$. Conversely, longitudinal fluctuations (along the $b$-axis) lead to a polarized signal for $\tau < 0$. Figure~\ref{fig:supp:CRO214_NSE} shows a significant polarization at $\tau < 0$ that depolarizes as a function of $\tau$ and $T$, respectively, indicative for the $T$-dependent linewidth $\Gamma(T)$ of the critical fluctuations. By contrast, no significant polarization at $\tau > 0$ was measured. Moreover, no oscillations of the polarization at small $\tau$ were observed, which are a hallmark of interference effects between transverse and longitudinal fluctuations \cite{Tseng.2016}. Hence, we conclude that transverse fluctuations along the $c$-axis are negligible or absent in \cro and we assume in the following that the fluctuations observed in the TAS experiments are also purely longitudinal. A possible explanation for this absence is that these fluctuations are gapped and thus not excited in the $T$-range of our study.} 

\begin{figure}[tb]
     \begin{center}
\includegraphics[width=0.85\columnwidth]{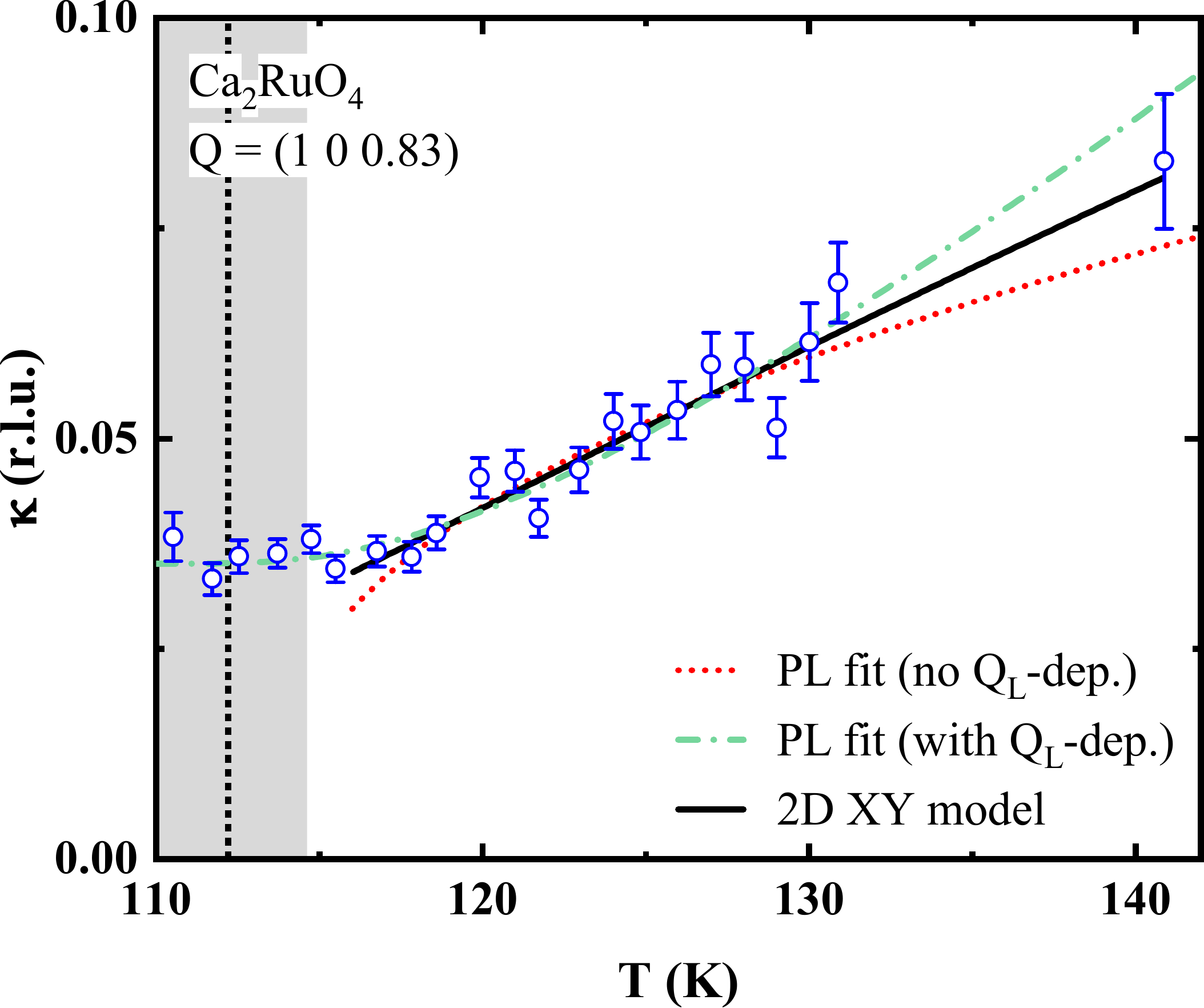}
         \caption{Inverse correlation length $\kappa(T)$ of \cro with 
various fit functions: The green dashed-dotted line is a PL fit with $\nu = 1.0(1)$ and $\Sub \kappa{3D} = \Sub QL\sqrt{J'/J} = 0.035(1)$. The red dashed line is a PL fit for $T > \SI{116}{\K}$ and $\kappa(T \leq \Sub TN) = 0$, with $\nu = 0.42(4)$.  The black solid line corresponds to a 2D-XY fit for $T > \SI{116}{\K}$, with $\Sub T{KT} = \SI{87(2)}{\K}$ and $b = 1.9$. The vertical dashed line 
shows the average \Sub TN and the grey bar the variance of \Sub TN.}
         \label{fig:results:CRO214_kappaT}
     \end{center}
\end{figure}

The inverse correlation length $\kappa(T)$ resulting from the fits in \fig{\ref{fig:results:QHscans_CRO214_examples}} is 
shown in \fig{\ref{fig:results:CRO214_kappaT}}. Notably, a $T$-dependent 
broadening above \SI{116}{\K} ($\Sub TN + \SI{4}{\K}$) can be observed, 
while the $Q$-width is approximately constant for $T < 116$\,K. This is 
in contrast to the conventional critical scaling theory where $\kappa$ 
should converge to zero at \Sub TN. In the following we will discuss 
several models to explain this saturation of $\kappa$ at $T < 116$\,K: 
(i) An obvious reason for such lower bound of the linewidth are 
crystallographic defects \cite{Birgeneau.1980,Cowley.1980}. One possible type of
defect in \cro can be domain walls of the structural twins, which can 
disrupt the long-range magnetic ordering. However, from the $Q$-width of 
the \peak100 magnetic peak (not shown here) we derive a domain size of $> 300\,\AA$, 
which is much larger than the extracted correlation length of $20\, \AA$ at $Q = \peak{1}{0}{0.83}$ and $T = 110\,K$. Thus we exclude domain size effects as the origin of the observed linewidth saturation. (ii) At $T \simeq \Sub TN$, \textit{i.e.} 
close to the 3D ordering, one expects a crossover of the critical fluctuations from a 2D to a 3D character with an increasing influence of 
the \Sub QL component on $\kappa$. This effect was described in 
Refs.~\cite{Hirakawa.1982,AlsNielsen.1993} and modelled by an effective 
$\kappa$ with $\Sub \kappa{eff}^2 = \Sub\kappa{3D}^2 + 
\Sub\kappa{pow}^2$, $\Sub\kappa{3D}^2 \equiv \Sub QL^2J'/J$, and 
$\Sub\kappa{pow}^2 \equiv \kappa_0 t^{2\nu}$. A fit of our experimental 
$\kappa(T)$ with a convolution of $\Sub \kappa{eff}$ and the 
aforementioned Gaussian distribution of $\Sub TN$ results in an exponent $\nu = 1.0(1)$ and $\kappa_{3D} = 0.035(1)$, and describes 
the data over the entire measured $T$-range (green dashed-dotted line in \fig{\ref{fig:results:CRO214_kappaT}}). The exponent $\nu$ matches 
the universal value of the 2D-I model ($\Sub \nu{2DI} = 1$ 
\cite{Collins.1989}). From $\Sub\kappa{3D} = 0.035(1)$ we obtain the ratio 
$J'/J = 0.002$. This is in agreement with $J'/J = 0.004$ derived from 
INS on 1\% Ti-doped \cro \cite{Kunkemoller.2015}. Assuming $J = 
\SI{5.8}{\milli\electronvolt}$ \cite{Jain.2017}, this corresponds to an interlayer coupling $J' \approx \SI{0.01}{\milli\electronvolt}$.

The red dashed line in \fig{\ref{fig:results:CRO214_kappaT}} shows a PL 
fit $\kappa \propto |t|^{\nu}$ without a $\kappa$-offset at \Sub TN, \textit{\ie} without a \Sub QL dependence due to 3D correlations, convoluted with the variance of \Sub TN. Only the data for $T > 
\SI{116}{\K}$, \ie beyond the saturation region, were included in the 
fit. The rounded shape of the red line towards \Sub TN results from the 
\Sub TN variance. The resulting $\nu = 0.42(4)$ is close to the 
mean-field (MF) value of $\Sub \nu{MF} = 0.5$, but is at odds with
$\nu = 1.0(1)$ obtained in the previous PL fit with the offset in 
$\kappa$, although both fits give a satisfactory description of the data 
for $T > 116\,K$. 
Furthermore, we note that a fit of $\kappa(T)$  with a 2D quantum Heisenberg model  \cite{Chakravarty.1988,Chakravarty.1989, Hasenfratz.1991} with an anisotropy parameter accounting for 3D correlations close to $\Sub TN$ 
\cite{Keimer.1992} also gives a good agreement with the data, but we 
exclude this model due to the large easy-plane anisotropy in \cro.

We now focus on the 2D-XY model, which was already suggested in the context of the magnon 
dispersion \cite{Jain.2017} and describes a topological phase transition accompanied by an unbinding of vortex/antivortex pairs \cite{Bereszinskii.1972,Kosterlitz.1973, 
Kosterlitz.1974}. The parameters of this model are the 
Kosterlitz-Thouless (KT) temperature $\Sub T{KT}$, a critical exponent $\eta = 0.25$, and a dimensionless non-universal parameter $b$
\cite{Kosterlitz.1974, Mertens.1989}, which was previously determined to be approximately 1.9 \cite{AlsNielsen.1993}. The correlation length in this 
model is defined as \cite{Kosterlitz.1974}
\begin{equation}
         \xi \propto \exp\left(\frac{b}{\sqrt{\Sub t{KT}}}\right), \quad 
\textrm{with } \Sub t{KT} \equiv (T/\Sub T{KT} - 1).
     \label{equ:models:2DXY_kappa}
\end{equation}
For systems with magnetic long-range order the actual KT-transition at $\Sub T{KT} < \Sub TN$ is usually obscured by the 3D ordering with nonzero interlayer couplings $J'$, which set in around \Sub TN.
The relation between $\Sub T{KT}$ and \Sub TN is given by 
\cite{Bramwell.1993, AlsNielsen.1993}
     \begin{equation}
         \frac{\Sub TN - \Sub T{KT}}{\Sub T{KT}} = 
\frac{4b^2}{[\ln(J/J')]^2}.
     \label{equ:models:TKT_TN}
     \end{equation}
Assuming $J'/J = 0.002$, as derived from the above PL fit with \Sub QL dependence to capture the $\kappa$-offset, we obtain 
$\Sub {\tilde{T}}{KT} = 82$\,K, with $\Sub {\tilde{T}}{KT}$ denoting the KT temperature derived from Eqn.~\eqref{equ:models:TKT_TN} and \Sub T{KT} the KT temperature extracted from fits to $\kappa(T)$ in the following. As expected for a system with a KT temperature below \Sub TN, our data do not show any signatures of a transition around 82\,K. Nonetheless, we use this model in the following to describe the scaling above \Sub TN, as it was demonstrated \cite{Roscilde.2003, Cuccoli.2003, Cuccoli.2003b} and 
experimentally confirmed \cite{Hirakawa.1982, L.P.Regnault.20, 
Heinrich.2003}, that even a XY anisotropy much weaker than in the case 
of \cro can result in 2D-XY scaling. Hence, as a next step, we fit Eqn.~\eqref{equ:models:2DXY_kappa} to $\kappa(T)$ for $T > \SI{116}{\K}$. Note that \Sub T{KT} is much lower than the lower limit of the fitting range, \ie a possible distribution of \Sub TN and KT temperatures in the sample will not affect the result of the fit significantly and is therefore not considered here. The resulting fit (solid black line in \fig{\ref{fig:results:CRO214_kappaT}}) with $\Sub T{KT} = \SI{87(2)}{\K} \approx 0.8 \Sub TN$ provides an excellent description of the data, and is in reasonable agreement with $\Sub {\tilde{T}}{KT} = 82$\,K from Eqn.~\eqref{equ:models:TKT_TN}. We note that a fit with $b$ as a free parameter did not converge, since it couples strongly to $\Sub T{KT}$. Thus, $b$ was fixed to 1.9 \cite{AlsNielsen.1993}. We also note that $\eta$ was fixed to 0.25 \cite{Kosterlitz.1973}, although  $\eta$ can deviate from this value in specific models on critical scattering, which then would affect the line-shape of $\X'(\Vb q)$ from the Lorentzian form employed in this work (see Methods).  However, we find that in our case the deviation of the line-shape for $\eta = 0.25$ is relatively subtle (see App.~\ref{app:PeakShape}) and lies below the detection threshold of the statistics of our data. Therefore, we use the simplest approach for $\X'(\Vb q)$ in the present work, which is the Lorentzian-function. 

In summary, the data in \fig{\ref{fig:results:CRO214_kappaT}} are consistent with both, the 2D-XY (black solid line) and 2D-I (green dashed-dotted line) scaling behavior for $T > 116$\,K. Nevertheless, we rule out the latter scaling for the description of $\kappa(T)$, as the critical peak amplitudes $S_0(T)/T$ [\fig{\ref{fig:results:CRO214_S0}}] with the corresponding critical exponent $\gamma$ (see below) are not compatible with the 2D-I model, although a crossover to 2D-I scaling close to \Sub TN is expected due to the orthorhombic anisotropy $\epsilon$ \cite{Jain.2017}. While this crossover from 2D-XY to 2D-I scaling presumably occurs in a $T$-range very close to \Sub TN and is not resolved in our data, we attribute the observed saturation for $T < 116$\,K to a crossover to 3D coupling, which eventually drives the magnetic transition.

In addition to $\kappa(T)$, we analyzed the staggered susceptibility $\X_0$, which also shows critical behavior close to \Sub TN. Related via the Kramers-Kronig relation, $\X_0$ is proportional to $S(0) \propto \X_0T$ (see Methods), \ie the peak amplitude of the Lorentzian-profile $S(q)$. In the following, $S(0)$ will be denoted as $S_0$. Figure~\ref{fig:results:CRO214_S0} shows the temperature dependence of the amplitude measured at \peak10{0.83}. First, we fit a PL $\propto 
|t|^{-\gamma}$ in the range $\SI{110}{\K} < T < \SI{140}{\K}$, convoluted with the Gaussian \Sub TN distribution by 
assuming $S_0 = 0$ for $T < \Sub TN$. The agreement with the data is not 
convincing and the extracted critical exponent $\gamma = 0.47(2)$ does 
not match universal values \cite{Collins.1989}, especially not the value predicted for the \mbox{2D-I} model ($\Sub \gamma{2DI} = 1.75$). Furthermore, since scaling theory predicts PL behavior for temperatures both above and below \Sub TN, we also carried out a fit over the entire $T$-range (see App.~\ref{app:Criticalamplitudes}). Nevertheless, such fit yields a similar value for $\gamma$, corroborating that PL scaling is not suitable to capture the temperature dependence of the amplitudes. Next, we fit the range $\SI{110}{\K} < T < \SI{140}{\K}$ with the 2D-XY model by using the scaling relation $\X_0 
\propto \xi^{2-\eta}$ \cite{Kosterlitz.1974}
     \begin{equation}
         \frac{S(0)}{T} \propto \exp\left(\frac{B}{\sqrt{\Sub t{KT}}}\right),
     \label{equ:models:2DXY_amplitude}
     \end{equation}
with $B \equiv b(2 - \eta)$ and $\Sub T{KT} = \SI{87}{\K}$ from above. We fixed $\eta = 0.25$ as suggested for the 2D-XY model \cite{Kosterlitz.1974}. The model gives a good description of the 
data with only one free parameter in the fit, that is, the proportionality constant in Eqn.~\eqref{equ:models:2DXY_amplitude}. 

\begin{figure}[tb]
     \begin{center}
\includegraphics[width=0.85\columnwidth]{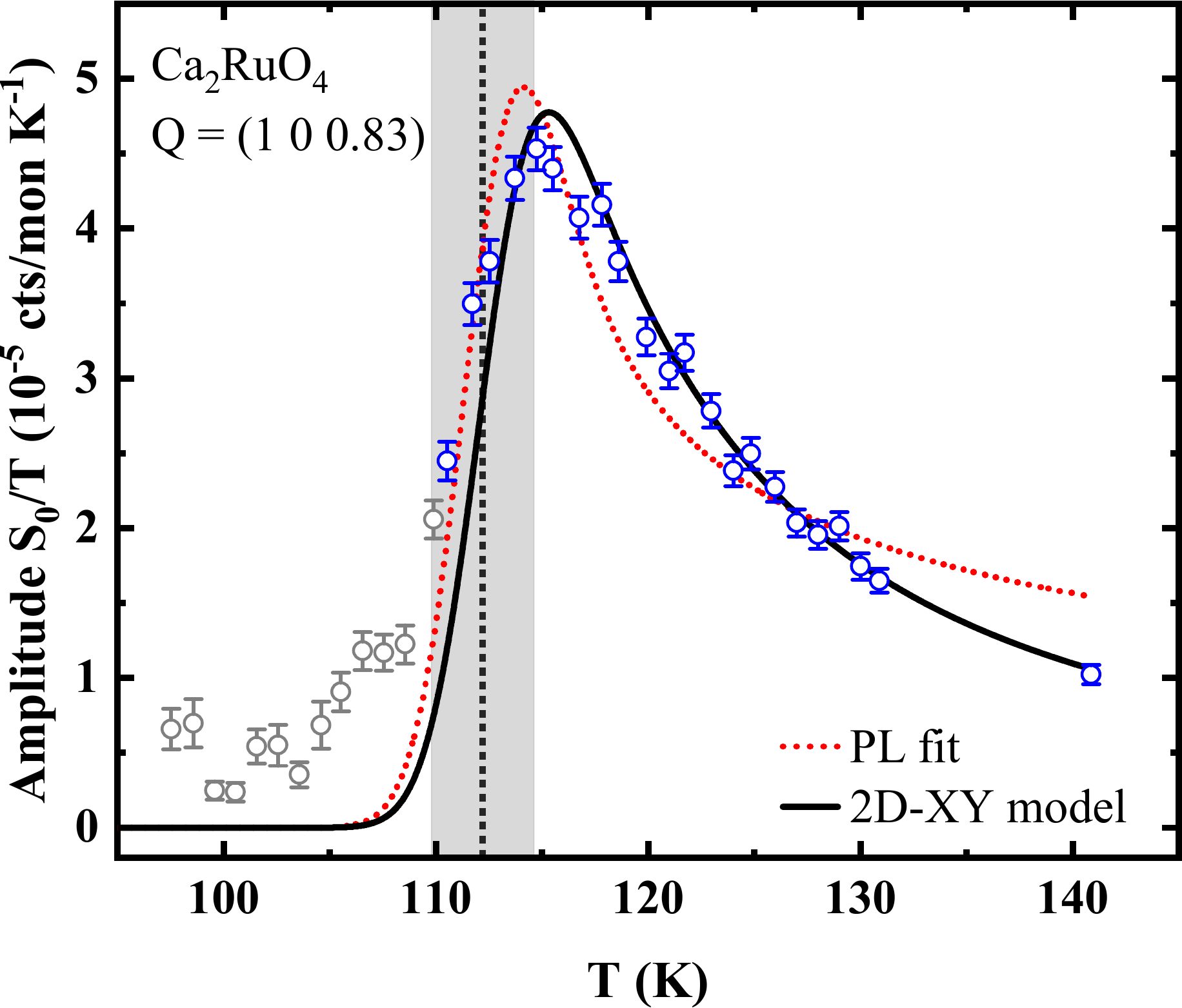}
         \caption{Peak amplitude $S_0(T)/T$ of \cro. The red dotted line is a PL scaling fit ($\gamma = 0.47(2)$) and the black solid line corresponds to the 2D-XY model. The grey data points 
were not included in the fit. The black vertical line indicates \Sub TN 
and the grey bar the variance of \Sub TN.}
         \label{fig:results:CRO214_S0}
     \end{center}
\end{figure}
\subsection{Static critical properties of \bicro}

\bicro exhibits ferromagnetic (FM) bilayers [\fig{\ref{fig:results:Tscan}}d], which are stacked in an AFM fashion along the $c$-axis (A-type AFM) \cite{McCall.2003,Yoshida.2005}. From the magnon dispersion in the ordered phase \cite{Bertinshaw.2021,Ke.2011b} the following terms of the Hamiltonian were derived: $J = \SI{-3.75}{\milli\electronvolt}$, $J_c = \SI{-6.5}{\milli\electronvolt}$ for the nearest neighbor and intra-bilayer coupling; and $E = \SI{5.5}{\milli\electronvolt}$, $\epsilon = \SI{2.5}{\milli\electronvolt}$ for the tetragonal and orthorhombic anisotropy, respectively. Notably, the magnon dispersion along the $c$-direction and inter-bilayer coupling $J'$ are very small or absent \cite{Bertinshaw.2021, Ke.2011b}. Hence, \bicro exhibits an easy-plane anisotropy $E$ and 
strongly coupled bilayers (large $J_c$) that can possibly act as one magnetic entity \cite{Ke.2011}, suggesting that \bicro could also be a candidate for quasi-2D-XY critical behavior. This calls for an investigation whether the critical behavior in the bilayer compound \bicro falls either into the quasi-2D or the 3D limit, or corresponds to an intermediate case.

Figure~{\ref{fig:results:Tscan}}e shows the magnetic \peak001 peak intensity \Sub I{001} measured upon warming, with a first-order transition at $\Sub T{N,2} \approx \SI{48}{\K}$ and a second-order transition at $\Sub T{N,1} \approx \SI{56}{\K}$, in good agreement with Ref.~\cite{Cao.1997d}.  
In contrast to the magnetic peak of \cro [\fig{\ref{fig:results:Tscan}}c], \Sub I{001} of \bicro drops sharply towards \Sub T{N,1}, suggesting that possible distribution of N\'{e}el temperatures $\Delta \Sub T{N,1}$ is negligible. We explain this observation with the appearance of less pronounced intrinsic crystal strains above the structural transition at \Sub T{N,2}. Moreover, in comparison to \cro, the crystal-field distortions are expected to be weaker in \bicro \cite{Bertinshaw.2021}. In order to establish the presence of critical scattering we measure the scattering intensity in distance to the \peak001 Bragg position at $Q = \peak{0}{0}{1.3}$ as a function of temperature (see inset \fig{\ref{fig:results:Tscan}e}). Notably, the critical intensity in the inset in \fig{\ref{fig:results:Tscan}}e peaks at \Sub T{N,1} and its magnitude is compatible with the remaining intensity for $T > \Sub T{N,1}$ in \fig{\ref{fig:results:Tscan}}e. A PL fit (without a $\Sub T{N,1}$ distribution) in the range $\SI{49}{\K} < T < \Sub T{N,1}$ ($0 < |t| < 0.1$), \ie between \Sub T{N,1} and \Sub T{N,2}, yields a critical exponent $\beta = 0.230(6)$ and \Sub T{N,1} = 54.16(2)\,K [\fig{\ref{fig:results:Tscan}}e,f]. Figure~\ref{fig:results:Tscan}f shows \Sub I{001} and the PL fit on a double-logarithmic scale, suggesting a purely linear evolution of \Sub I{001} in such a plot for the measured temperatures. 
The obtained value of $\beta$ matches the universal value of the 2D-XY model ($\beta = 0.23$ \cite{Bramwell.1993,Bramwell.1993b}), although it should be taken with caution as the point density in close vicinity of \Sub T{N,1} is sparse. Moreover, the intrinsic scaling behavior could be obfuscated due to a contribution in the scattering intensity from an overlap with the second transition at \Sub T{N,2}. 

\begin{figure}[htbp]
	\begin{center}
	\includegraphics[width=\columnwidth]{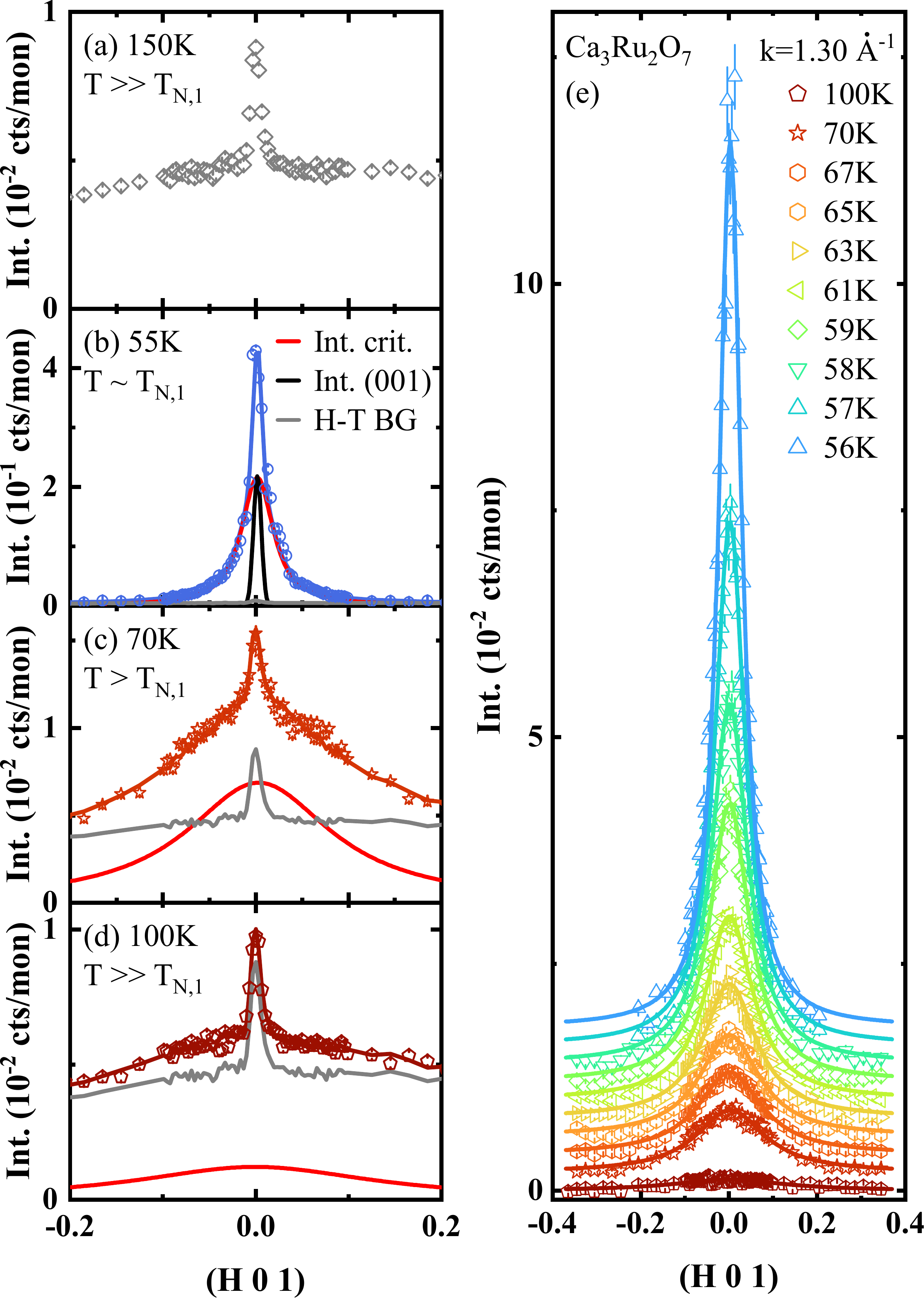}
		\caption{Selected energy-integrated transverse \Sub QH-scans around \peak H01 for \bicro before (a-d) and after (e) BG subtraction. (a) The high-$T$ BG contains a sharp peak that is independent of \Sub QL. (b) Close to $\Sub T{N,1}$, a resolution limited Gaussian-component arising from the \peak001 magnetic Bragg-peak is observed in addition to the Voigt-profile of the critical scattering. (c,d) Above $\Sub T{N,1}$, the sharp peak and critical scattering are observed. (e) Selected \Sub QH-scans after BG subtraction with corresponding fits (solid lines). For clarity the data are plotted with a constant offset.}
		\label{fig:results:QHscans_CRO327_examples}
	\end{center}
\end{figure}

To extract $\kappa(T)$ and the amplitude of the critical scattering, we carried out \Sub QH-scans around \peak{H}01 in the two-axis mode. The ideal energy-integrating configuration used for \cro is not applicable for 3D systems (see Methods). However, a numerical simulation confirmed that the integration according to Eqn.~\eqref{equ:ScatTheo:approxBirg} is sufficiently satisfied for our two-axis configuration in \bicro (see App.~\ref{app:energyIntegrationBicro}). \added{Furthermore, we note that for \bicro, spin-echo experiments to discern longitudinal and transverse fluctuations have not been carried out. Nonetheless, from the fact that in the present study both, the static and dynamical critical fluctuations in \bicro are well captured by power-laws (see below), we conclude that the non-critical transverse fluctuations do not contribute significant intensity around \Sub T{N,1} \cite{Dietrich.1969}.} 
From the \Sub QH-scans we subtracted two BG components: (i) A sharp peak at $H=0$, which is clearly visible at high-$T$ (150\,K)  [\fig{\ref{fig:results:QHscans_CRO327_examples}}a]. In addition, we performed a scan around \peak{H}0{1.25} at 100\,K and found that the sharp feature is independent of \Sub QL and $T$ (see App.~\ref{app:sharpPeakBicro}). Thus, we assign it to 2D diffuse nuclear scattering from disorder along the $c$-axis due to \eg stacking faults. (ii) The sharp resolution-limited \peak001 peak with Gaussian-width $\approx 0.01$\,r.l.u. [\fig{\ref{fig:results:QHscans_CRO327_examples}}b], which rapidly vanishes above \Sub T{N,1}. The \peak001 peak is intense at $T \leq 54.5$\,K and the extraction of critical scattering is not reliable. The BG corrected \Sub QH-scans can be captured by a Voigt-profile (intrinsic Lorentzian critical scattering convoluted with instrumental Gaussian-profile) [\fig{\ref{fig:results:QHscans_CRO327_examples}}e]. A critical scattering intensity can be clearly observed at least up to 70\,K. We note that there might be a contribution from critical fluctuations even at 100\,K, but since it is very weak we do not consider this temperature in the following analysis.

\begin{figure}[tb]
	\begin{center}
	\includegraphics[width=0.85\columnwidth]{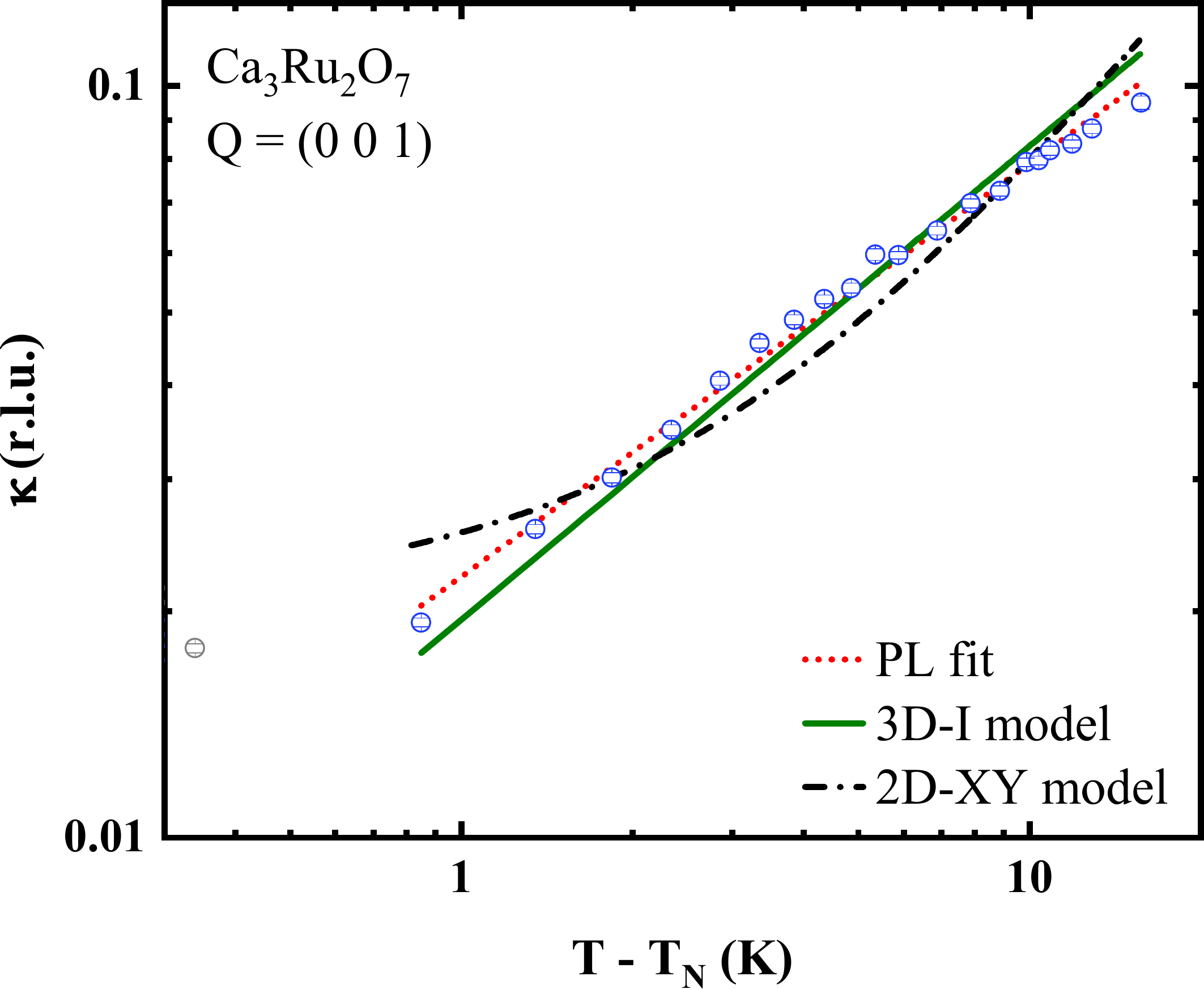}
		\caption{Inverse correlation length $\kappa(T)$ of \bicro on double logarithmic scales. The red dotted line is a linear fit in the range $\SI{55}{\K} \leq T \leq 70\,K$, with the slope corresponding to the critical exponent $\nu = 0.550(4)$, according to the scaling relation $\kappa \propto t^\nu$. The green solid line indicates 3D-I scaling. The black dashed-dotted line corresponds to the 2D-XY model with $\Sub T{KT} = 45.42(6)$\,K and $b = 1.9$. The grey data point was not included in the fit.}
		\label{fig:results:CRO_kappaT_log}
	\end{center}
\end{figure}

The resulting $Q$-width $\kappa(T)$ is plotted in \fig{\ref{fig:results:CRO_kappaT_log}} on double logarithmic scales. Note that in the following analysis of \bicro, we extract the critical exponents of the PLs from the slopes of linear fits in plots with double logarithmic scaling, whereas plots with linear scaling were employed in the above analysis of \cro  [\fig{\ref{fig:results:CRO214_kappaT}} and \fig{\ref{fig:results:CRO214_S0}], due to the variance of \Sub TN in the latter material.} The red dotted line is a linear fit in the range $\SI{55}{\K} \leq T \leq \SI{70}{\K}$ with the slope corresponding to the critical exponent in the scaling relation $\kappa \propto t^{\nu}$. The obtained $\nu = 0.550(4)$ lies between the values predicted for the 3D-I ($\Sub \nu{3DI} = 0.630$, \cite{Pelissetto.2002}) and the MF model ($\Sub \nu{MF} = 0.5$, \cite{Collins.1989}). The MF model, however, is at odds with the modeling of the magnon dispersion of \bicro in Ref.~\cite{Bertinshaw.2021}, which used only nearest neighbor couplings and no long-ranged interactions in the spin Hamiltonian \cite{Collins.1989}. Thus, we assign the critical scaling of $\kappa(T)$ rather to the 3D-I model (green solid line in \fig{\ref{fig:results:CRO_kappaT_log}}), which also captures the data well. 
In analogy to \cro, we also fit $\kappa(T)$ of \bicro with the 2D-XY model, using Eqn.~\eqref{equ:models:2DXY_kappa}, $\eta = 0.25$, and $b = 1.9$. The obtained KT-temperature is $\Sub T{KT} = 45.42(6)$\,K, but the agreement between the fit (black dashed-dotted) and the data is unsatisfactory for most temperatures [\fig{\ref{fig:results:CRO_kappaT_log}}]. We also test a fit with $b$ as a free parameter, since lower values of $b$ were reported in some experiments \cite{Gaveau.1991,Hemmida.2009} and derived in numerical calculations \cite{Mertens.1989}. Such a fit (not shown here), with $b = 0.44(1)$ and $\Sub T{KT} = 53.18(7)$, yields a better agreement with the $\kappa(T)$ data of \bicro. However, using the latter values of $b$ and \Sub T{KT} as an input for the fit of the critical amplitudes (see below) results in a very strong deviation from the data (not shown here) and is therefore disregarded. In summary, we conclude that the 3D-I model is most appropriate to describe the critical behavior of $\kappa(T)$ above \Sub TN. 

\begin{figure}[tb]
	\begin{center}
	\includegraphics[width=0.85\columnwidth]{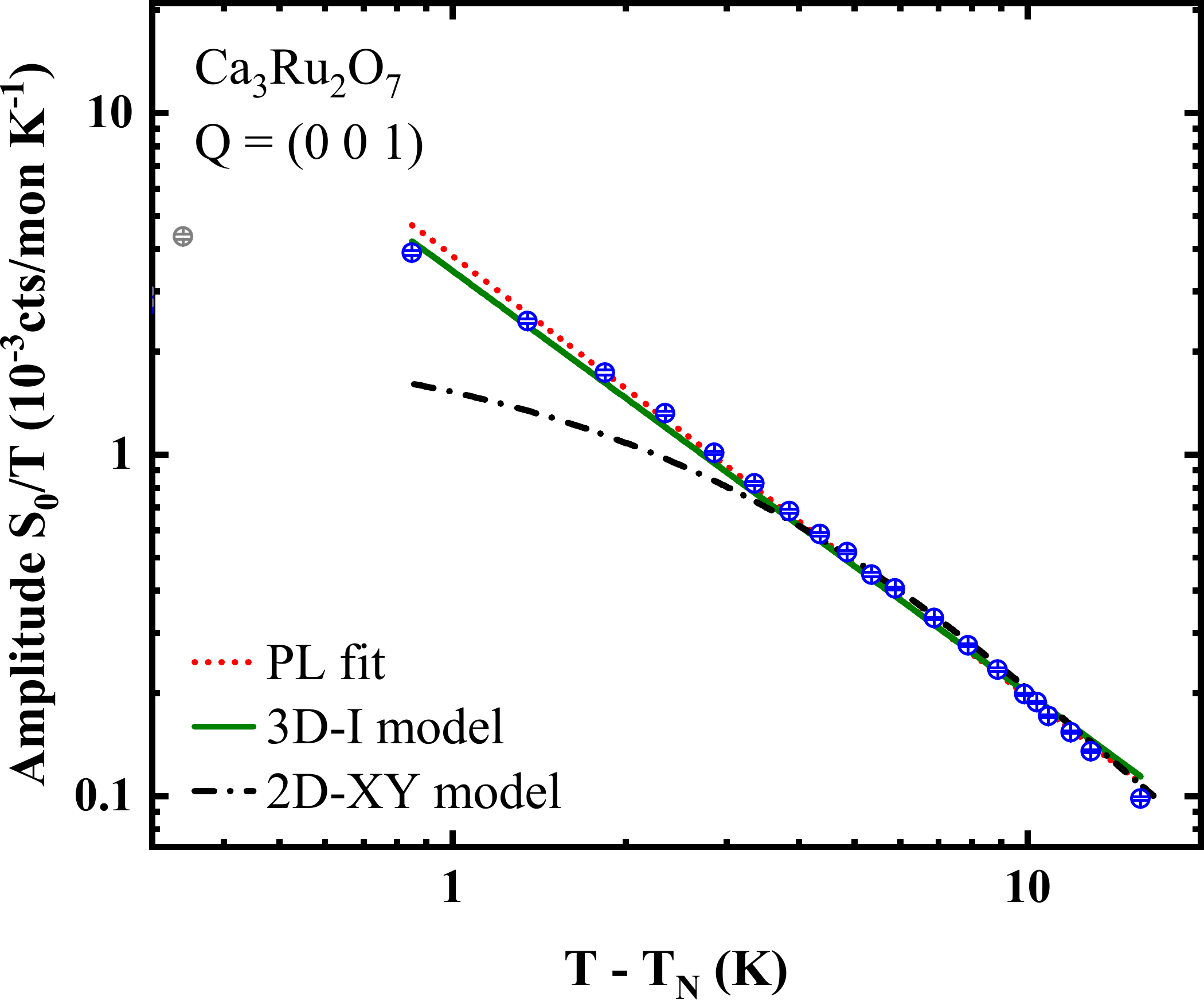}%v4
		\caption{Peak amplitude $S_0(T)/T$ of \bicro on double logarithmic scales. The red dotted line is a linear fit with the slope corresponding to the critical exponent $\gamma = 1.290(4)$, according to the scaling relation $\X \propto t^{-\gamma}$. The green solid line indicates 3D-I scaling. The black dashed-dotted line corresponds to the 2D-XY model with $\Sub T{KT} = 45.42(6)$\,K and $b = 1.9$. The grey data point was not included in the fit.}
		\label{fig:results:CRO_S0_log}
	\end{center}
\end{figure}

The critical exponent $\gamma$ of the staggered susceptibility is obtained by fitting the corresponding peak amplitudes $S_0/T$ with the PL scaling $\X \propto t^{-\gamma}$ in the range $\SI{55}{\K} \leq T \le \SI{70}{\K}$. From the slope of the corresponding linear fit (red dotted line) in the double-logarithmic plot [\fig{\ref{fig:results:CRO_S0_log}}], we extract $\gamma = 1.290(4)$, which is close to the value predicted for the 3D-I model (\Sub \gamma{3DI} = 1.238, \cite{Pelissetto.2002}), as indicated by the green solid line in \fig{\ref{fig:results:CRO_S0_log}}. For the comparison with the 2D-XY model, we use Eqn.~\eqref{equ:models:2DXY_amplitude}, with $\eta = 0.25$ and $b = 1.9$, as well as $\Sub T{KT} = 45.42$\,K determined from the fit of $\kappa(T)$ above. In spite of a good agreement with the data at $T > \Sub TN + 2$\,K (see black dashed-dotted line), the PL fits are more suitable to describe the scaling of the critical amplitudes closer to \Sub TN.

\subsection{Dynamic critical properties of \cro}

\begin{figure}[tb]
	\begin{center}
	\includegraphics[width=\columnwidth]{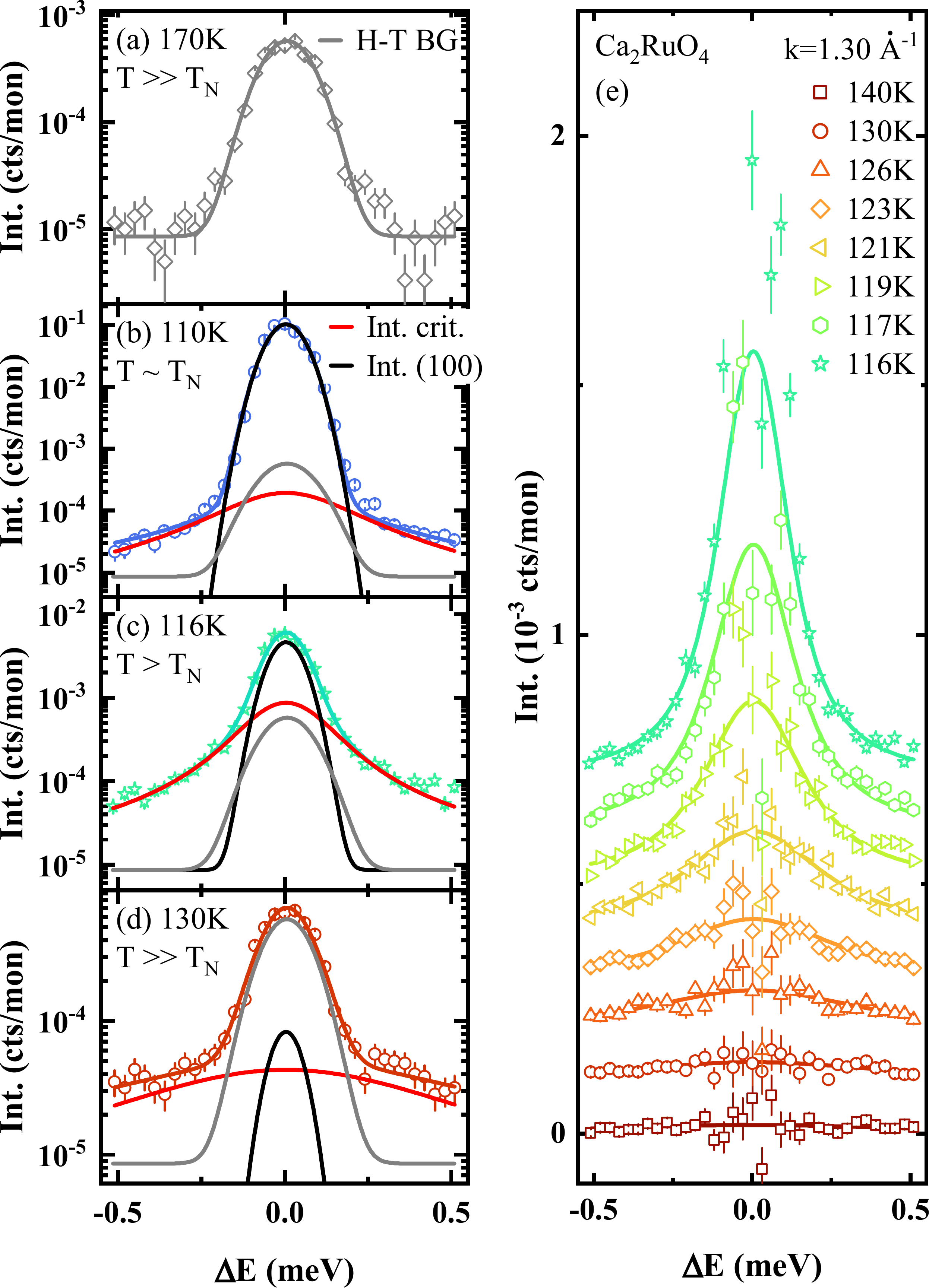}
		\caption{Selected energy scans of \cro before (a-d) and after (e) BG subtraction. (a-d) Besides a constant high-$T$ BG (170\,K), we fitted a sum of a broad critical Voigt-profile and a resolution limited Gaussian-peak over the entire T-range. The latter is attributed to the magnetic \peak100 peak that shows the aforementioned distribution of \Sub TN. (e) Selected energy scans after high-$T$ BG and elastic \peak100 peak subtraction with corresponding fits (solid lines). For clarity the curves are plotted with a constant offset.}
		\label{fig:results:CRO214_Eall}
	\end{center}
\end{figure}

Figures~\ref{fig:results:CRO214_Eall}a-d display selected energy-scans at \peak100 measured on the cold neutron TAS FLEXX with $k_f = \kvec{1.3}$. A $T$-independent BG from elastic incoherent scattering recorded at 170\,K was subtracted from the raw data. The BG corrected data are described by a fit with the sum of a resolution-limited Gaussian-peak (FWHM $\approx 0.12$\,meV) and a Voigt-function capturing the critical scattering. The resolution was extracted from the width of the elastic magnetic scattering well-below \Sub TN. The free parameters of the fits are the amplitude of the Gaussian-peak, and the amplitude and width of the Voigt-peak. The amplitudes of the Gaussian and the Voigt match the intensities of the \peak100 peak [\fig{\ref{fig:results:Tscan}}b] and the amplitude of the critical scattering at \peak10{0.83} [\fig{\ref{fig:results:Tscan}}b, inset] (see App.~\ref{app:EscanIntensitiesCro}).
An overview of the energy-scans and the resulting fit curves at selected temperatures is shown in \fig{\ref{fig:results:CRO214_Eall}e}. 

\begin{figure}[tb]
	\begin{center}
	\includegraphics[width=0.85\columnwidth]{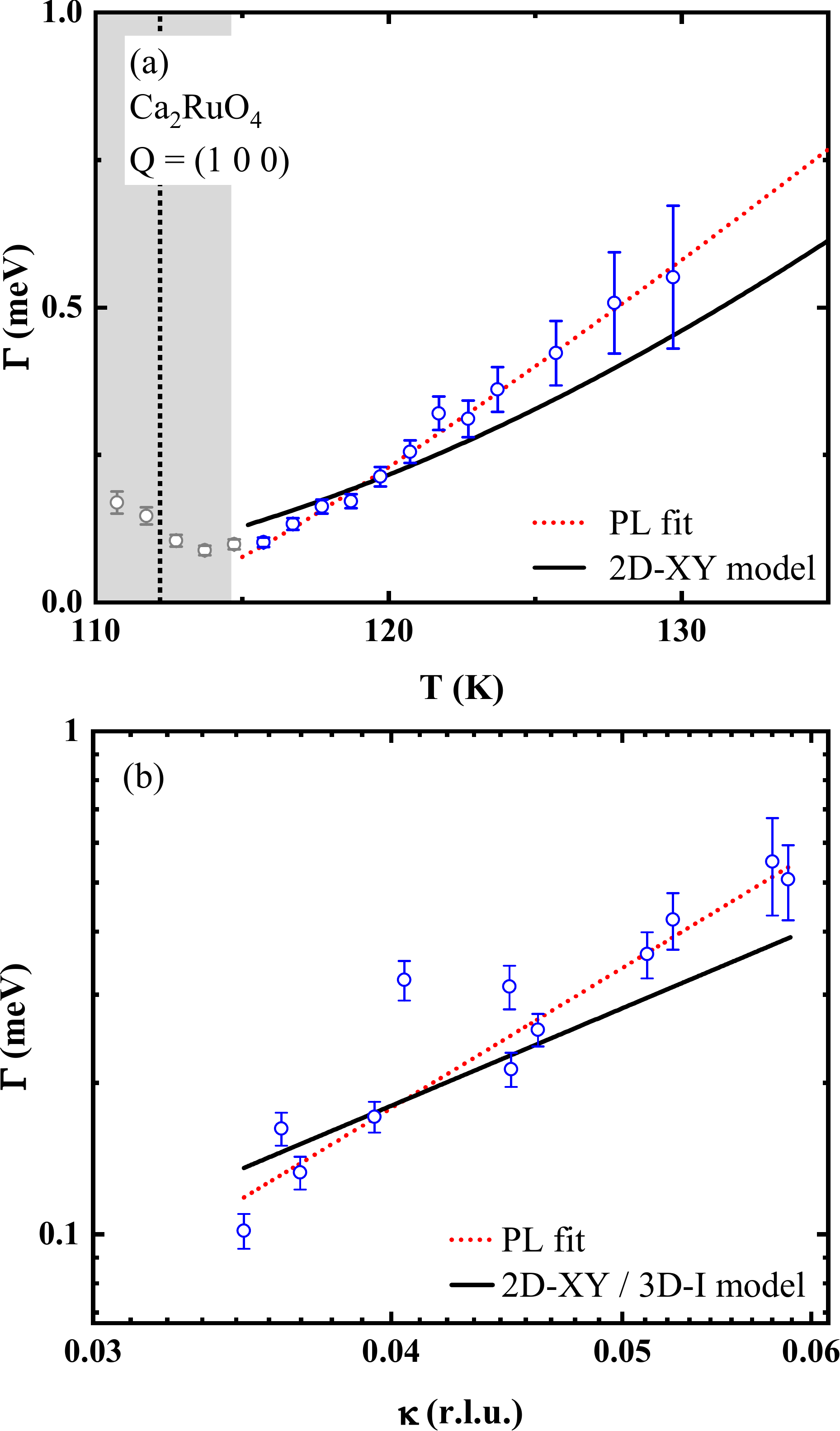}
		\caption{(a) Energy-width $\Gamma$ of \cro \vs temperature. The red dotted line is a PL scaling fit with exponent $z\nu = 1.1(1)$. The black solid line corresponds to a 2D-XY scaling fit with $\Sub z{XY} = 2.0$ \cite{Jensen.2000}. The grey data points are not included in the fits. The black vertical line indicates \Sub TN and the grey bar the variance of \Sub TN. (b) Energy-width $\Gamma$ of \cro \vs the inverse correlation length $\kappa$ on double-logarithmic scales. The red dotted line is a linear fit with the slope corresponding to the dynamical critical exponent $z = 2.9(2)$, according to the scaling relation $\Gamma \propto \kappa^z$. The black solid line corresponds to the 2D-XY model with $\Sub z{XY} = 2.0$ \cite{Jensen.2000}. 
		}
		\label{fig:results:CRO214_kappaGamma}
	\end{center}
\end{figure}

    The resulting energy-width $\Gamma(T)$ of the critical component is plotted in \fig{\ref{fig:results:CRO214_kappaGamma}a}. Above \SI{115}{\K} the data show a significant broadening, while below no systematic trend was observed. For these data ($T \leq \SI{115}{\K}$) the intensity of the \peak100 peak is much stronger than the critical scattering and extraction of $\Gamma(T)$ in the fit is not reliable. Therefore, we exclude this $T$-range in the following scaling analysis (grey data points in \fig{\ref{fig:results:CRO214_kappaGamma}a}). To determine the dynamic critical exponent $z$ from $\Gamma(T)$ above $\SI{115}{\K}$, in analogy to $\kappa(T)$, we fit a convolution of the PL $\Gamma \propto |t|^{z\nu}$ with a Gaussian distribution of $\Sub TN$ (red dotted line). By assuming $\Gamma(T) = 0$ for $T \leq \Sub TN$, as predicted by dynamic scaling theory \cite{Hohenberg.1977}, the PL fit yields $z\nu = 1.1(1)$. In spite of the good agreement with the data in \fig{\ref{fig:results:CRO214_kappaGamma}a}, the error margin is relatively large and only the 3D-H model ($z\Sub\nu{3DH} = 1.067$, \cite{Hohenberg.1977, Campostrini.2002}) and the 3D-I model ($z\Sub\nu{3DI} = 1.26$, \cite{Hohenberg.1977, Pelissetto.2002}) are reasonably close to the obtained exponent. However, 3D-scaling is not expected well above $\Sub TN$ for this quasi-2D system. Furthermore, $z\nu = 1.1(1)$ is far away from 2D-I scaling ($z\Sub\nu{2DI} = 1.75$ \cite{Mazenko.1981, Collins.1989}), which was discussed in the context of $\kappa(T)$ of \cro (green dashed-dotted line in \fig{\ref{fig:results:CRO214_kappaT}}).

Next, we examine the critical dynamics of \cro in terms of the 2D-XY model. For the motion of vortices a dynamic scattering function $S(\Vb q, \Vb \omega)$ with a quadratic Lorentzian form (central peak) was derived \cite{Huber.1982, Mertens.1987, Mertens.1989}, and also experimentally observed \cite{Hutchings.1986, Regnault.1986, L.P.Regnault.20}. However, such a central peak is not present in our data [\fig{\ref{fig:results:CRO214_Eall}}], which can be  captured by a simple Lorentzian function. Thus, we use instead
the dynamic scaling relation $\Gamma \propto \kappa^z$, suggested to be appropriate for relaxational dynamics in the 2D-XY model \cite{Jensen.2000} to check for 2D-XY scaling above $\SI{115}{\K}$. The black solid line in \fig{\ref{fig:results:CRO214_kappaGamma}a} shows the scaling for the critical exponent $\Sub z{XY} = 2.0$, which was postulated for the 2D-XY model \cite{Jensen.2000}. The agreement with the data is not as good as for the PL fit, although at high temperatures the 2D-XY scaling lies essentially within the error bars. The deviation at low temperatures could be due to the variance of \Sub TN or a crossover to a different scaling behavior. We note that using $z$ as a free fit parameter improves the agreement with the data, but the obtained value of  $\Sub z{XY} = 3.04(6)$ (fit not shown here) is at odds with the theory for the universality class. 

In addition, we determine $z$ directly from plotting $\Gamma$ \vs $\kappa$ on double logarithmic scales [\fig{\ref{fig:results:CRO214_kappaGamma}b}]. In such a plot, the slope of a linear fit to the data corresponds to the critical exponent of the scaling relation $\Gamma = \kappa^z$. The resulting dynamical critical exponent $z = 2.9(2)$ (red dashed line) is not compatible with any known universal value. Note, however, that the $\Gamma$ \vs $\kappa$ data points in \fig{\ref{fig:results:CRO214_kappaGamma}b} were obtained by interpolation, as the $Q$- and energy-scans on \cro were taken at different $Q$-positions and temperatures at two different instruments, which can lead to uncertainties. Specifically, the data points in \fig{\ref{fig:results:CRO214_kappaGamma}b} were generated by interpolating  $\kappa(T)$ to the temperatures at which $\Gamma(T)$ was measured and the same fitting range ($T > 115$\,K) was included, where the impact on the \Sub TN distribution is negligible. Nevertheless, while the dynamical critical exponent of the linear fit deviates from a universal value, we find that the 2D-XY model (solid black line) with the exponent $z = 2.0$ \cite{Jensen.2000} captures the $\Gamma$ \vs $\kappa$ data reasonably well [\fig{\ref{fig:results:CRO214_kappaGamma}b}].

\subsection{Dynamic critical properties of \bicro}

\begin{figure}[tb]
	\begin{center}
	\includegraphics[width=\columnwidth]{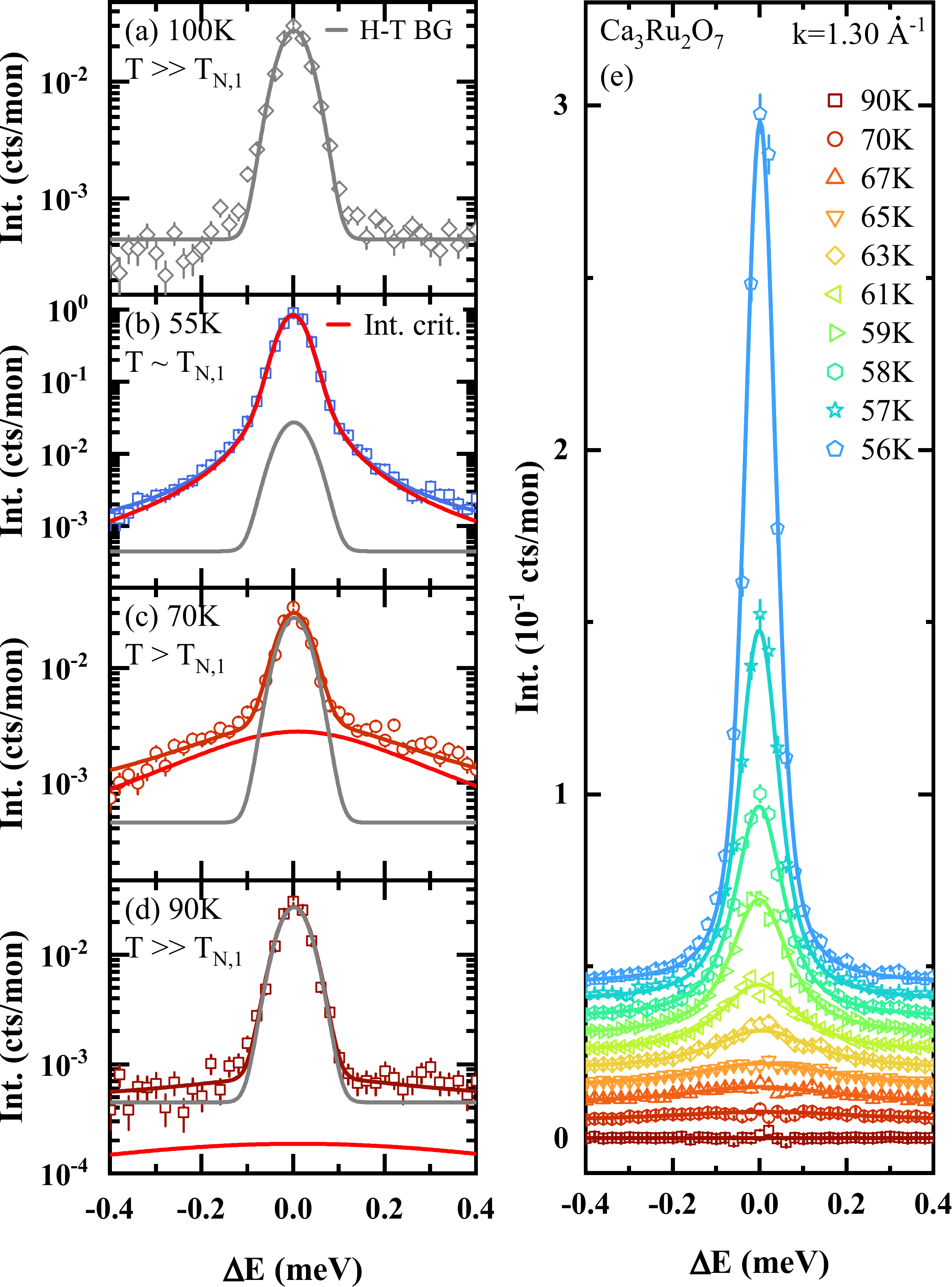}
		\caption{Selected energy scans of \bicro before (a-d) and after (e) BG subtraction. (a-d) In contrast to \cro, the \bicro data could be well fitted by a single Voigt-profile and a constant  elastic high-T BG measured at 100\,K. (d) Selected energy scans after BG subtraction with corresponding fit functions (solid lines). For clarity the curves are plotted with a constant offset.}
		\label{fig:results:CRO327_Eall}
	\end{center}
\end{figure}

Figures~\ref{fig:results:CRO327_Eall}a-d display selected energy-scans at the \peak001 peak of \bicro. A constant elastic incoherent high-$T$ BG is subtracted from the energy-scans, which was measured at $T = 100$\,K, where the contribution of critical scattering is negligible [\fig{\ref{fig:results:QHscans_CRO327_examples}d}]. Subsequently, a single Voigt-function (Gaussian-width $\approx 0.06$\,meV) is fitted to the scans. An overview of the energy-scans after BG subtraction with the corresponding fits is plotted in \fig{\ref{fig:results:CRO327_Eall}e} for selected temperatures.

\begin{figure}[tb]
	\begin{center}
	\includegraphics[width=0.85\columnwidth]{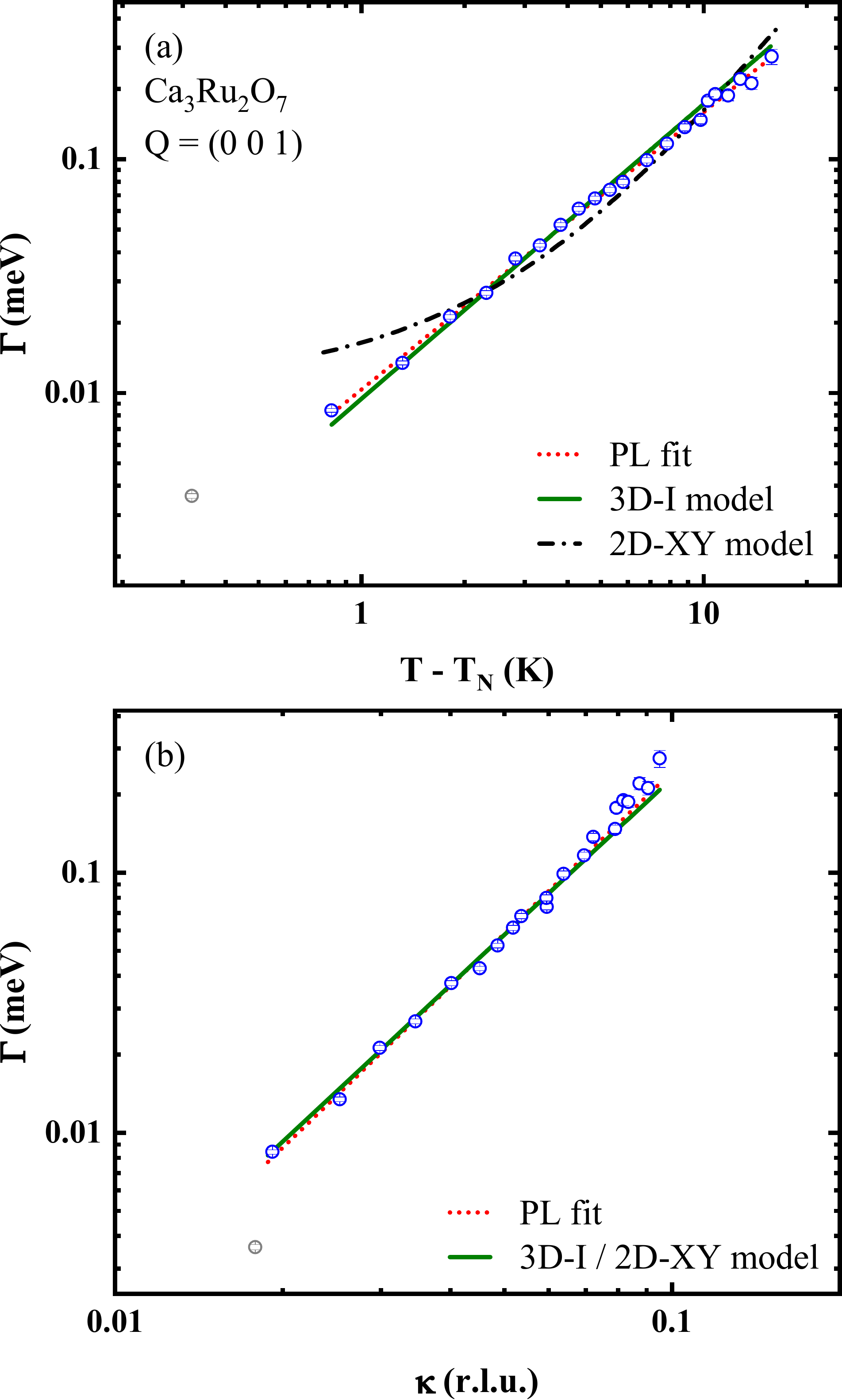}
		\caption{Energy-width $\Gamma$ of \bicro \vs temperature (a) and the inverse correlation length (b) on double-logarithmic scales. (a) The red dotted line is a linear fit with the slope corresponding to the critical exponent $z\nu = 1.186(8)$, according to the scaling relation $\Gamma \propto t^{z\nu}$. The green solid line indicates 3D-I scaling with $z\Sub \nu{3DI} = 1.260$ \cite{Hohenberg.1977, Pelissetto.2002}. The black dashed-dotted line corresponds to the 2D-XY model with the exponent $z = 2.0$ \cite{Jensen.2000}. (b) The red dotted line is a linear fit with the slope corresponding to the critical exponent $z = 2.14(2)$, according to the scaling relation $\Gamma \propto \kappa^z$. The green solid line corresponds to 3D-I scaling and the 2D-XY model, which both exhibit the exponent $z = 2.0$ \cite{Hohenberg.1977, Hasenbusch.2020,Jensen.2000}. Note that only the 3D-I model is also consistent with the static critical properties of \bicro. The grey data points were not included in the fits.
		}
		\label{fig:results:CRO327_kappaGamma}
	\end{center}
\end{figure}

The resulting energy-width $\Gamma(T)$ of \bicro is shown in \fig{\ref{fig:results:CRO327_kappaGamma}a} on double logarithmic scales. The data are well-captured by a linear fit (red dotted line) in the range $\SI{55}{\K} \leq T \le \SI{70}{\K}$ with a slope $z\nu = 1.186(8)$, corresponding to the critical exponent in the scaling relation $\Gamma \propto t^{z\nu}$. This exponent is close to the value predicted for the 3D-I model ($z\Sub \nu{3DI} = 1.260$, \cite{Hohenberg.1977, Pelissetto.2002}) indicated by the green solid line in \fig{\ref{fig:results:CRO327_kappaGamma}a}, which is consistent with the static critical properties. 
In analogy to \cro, we also carry out a fit with the 2D-XY model, using the dynamic scaling relation $\Gamma \propto \kappa^z$ with $z = 2.0$ \cite{Jensen.2000}. The resulting fit (black dashed-dotted line in \fig{\ref{fig:results:CRO327_kappaGamma}a}) describes the data reasonably well at high temperatures, but deviates strongly in proximity to \Sub TN.  

Figure~\ref{fig:results:CRO327_kappaGamma}b shows a plot of $\Gamma$ \vs $\kappa$ on double logarithmic scales, which allows us to determine $z$ directly. From the slope of a linear fit (red dotted line), we derive $z = 2.14(2)$ as the critical exponent, which is close to $z = 2.0$ proposed for the 3D-I model \cite{Hohenberg.1977, Hasenbusch.2020} and consistent with the $\Gamma(T)$ scaling above. Notably, $z = 2.0$ also corresponds to the 2D-XY model \cite{Jensen.2000}. However, since the 3D-I scaling is also compatible with the temperature-dependence of the $Q$-width and critical amplitudes at $T > \Sub TN$ of \bicro, we consider this model as most appropriate.

%-------------------------------------------
% DISCUSSION

\section{Discussion and conclusion}

For \cro, we revealed that the description of the static critical properties by PL scaling gave only partially satisfactory results, whereas application of the 2D-XY model provided a conclusive picture. In more detail, the obtained critical exponent $\beta = 0.158(6)$ is consistent with a 2D-XY model with fourfold crystal anisotropy (XY\textit{h}$_4$) \cite{Taroni.2008}. The inverse correlation length $\kappa$, in principle, could be fitted with different PLs and the 2D-XY model. Specifically, the observed saturation of $\kappa$ in the range between \Sub TN and \Sub TN + 4\,K was captured by a PL with \Sub QL-dependence, likely indicating the presence of 3D fluctuations close to \Sub TN. Nevertheless, the employed PL fits, with and without $\kappa$ offset, were not consistent with the 2D nature of \cro far above \Sub TN and results from fits of the amplitudes of the critical scattering.
Instead, the 2D-XY model captured both, $\kappa$ and the amplitude adequately. The resulting Kosterlitz-Thouless temperature $\Sub T{KT} = \SI{87(2)}{\K}$ was consistent with the ratio $J'/J = 0.002$ derived from the offset in $\kappa(T)$. Our attempt to describe the amplitude of the critical scattering by PL scaling showed that the functional form does not capture the data appropriately and the extracted critical exponent $\gamma = 0.47(2)$ is not compatible with universal values. Yet, 2D-XY scaling captured the critical amplitudes in a broad T-range around \Sub TN, in agreement with the Hamiltonian extracted from the magnon dispersion in the ordered phase \cite{Jain.2017}. 
The energy-width $\Gamma(T)$ is best captured by a PL fit with $z\nu = 1.1(1)$, which is compatible with 3D AFM Heisenberg scaling ($z\Sub \nu{3DH} = 1.067$, \cite{Hohenberg.1977, Campostrini.2002}). However, such a model for the dynamical critical scaling would be in stark contrast to the static critical behaviors, the magnon dispersion \cite{Jain.2017}, and the quasi-2D character of \cro. Consequently, we fitted the data with the 2D-XY model and $\Sub z{XY} = 2.0$ \cite{Jensen.2000}, which also captures the data reasonably well. 

In general, a crossover in the critical behavior from 2D-XY to Ising scaling close to \Sub TN can be expected in \cro due to the orthorhombic terms in the spin Hamiltonian \cite{Jain.2017}. The observed saturation of $\kappa(T)$ close to \Sub TN might be indicative for a dimensionality crossover to 3D scaling, whereas signatures of such a crossover were less clear in the analysis of the other critical exponents. 
This apparent absence can be due to the limited instrumental resolution of our TAS measurements and calls for a complementary high-resolution NSE study, focusing on the temperature range in close vicinity to \Sub TN. Notably, the enhanced energy resolution of NSE previously helped to resolve controversies about the scaling behavior of heavy Fermion superconductors \cite{Haslbeck.2019} and revealed a crossover from Heisenberg to Ising scaling close to \Sub TN in the classical 3D AFM MnF$_2$ due to uniaxial anisotropy \cite{Tseng.2016}. Moreover, subtle signatures of additional phases, which were proposed to exist in \cro above the AFM order, such as  orbital order \cite{Zegkinoglou.2005, Lotze.2021} and a Jahn-Teller driven spin-nematic phase \cite{Liu.2019}, might be detectable in the critical scaling behavior measured with high-resolution NSE. Apart from that, it will be interesting to probe the existence of possible vortex/antivortex-pairs with cryogenic microscopy techniques, such as Lorentz transmission electron microscopy \cite{Togawa.2021}, which might be particularly pronounced in thin films of \cro \cite{Dietl.2018}.

Overall, the distribution in \Sub TN in our sample introduces some uncertainty in our analysis of the critical scattering in \cro. Nonetheless, we find that our determination of the critical exponents of the $Q$-width, critical amplitudes, and energy-width is relatively insensitive to the details of the variance of \Sub TN. Along these lines, we performed fits (not shown here) assuming a difference of $\pm 0.5$\,K to our above value of $\Delta\Sub TN = \SI{4.84(1)}{\K}$. The resulting values for $\nu$ and $z$ are closely similar to the above ones (difference smaller than the error bars). This is plausible since we fit the data only for $T > \Sub TN + 4$\,K, where the impact of the \Sub TN distribution is relatively small. For the critical amplitudes, where the impact of the \Sub TN distribution on the critical exponent is expected to be strongest, as we fit in the range 110-140\,K, we find a deviation of only four percent. Also in the case of the critical amplitudes, the change of the critical exponent is smaller than the error bars. This result suggests that even a putative uncertainty in our value of $\Delta\Sub TN$ would not critically affect our determination of the critical exponents, corroborating the robustness of our analysis. A definitive determination of the critical behavior close to \Sub TN will require the synthesis of large, monolithic \cro single crystals, which appears to be out of reach of the methodologies currently at hand.

For the bilayer compound \bicro, which exhibits strong intra-bilayer couplings, the critical scaling was only partly compatible with the 2D-XY model. Although $\beta = 0.230(6)$ extracted from the temperature-dependence of the magnetic \peak001 peak matches the expected value for realistic 2D-XY systems \cite{Bramwell.1993,Bramwell.1993b}, the result should be taken with caution, due to a relatively low point density around \Sub T{N,1} and a possible overlap with the signal from the transition at \Sub T{N,2}. While the extracted $\beta$ is seemingly far from the corresponding value of the 3D-I model ($\Sub \beta{3DI} = 0.327$, \cite{Pelissetto.2002}), previous works in the context of Sr$_3$Ir$_2$O$_7$ pointed out that a significant underestimation of $\beta$ can arise when the power law analysis is not narrowly focused around \Sub TN  \cite{Vale.2019}, which provides a possible reconciliation between our small $\beta$  and the proposed 3D-I scaling.
The critical scaling of the $Q$-width, amplitude, and energy-width above \Sub T{N,1} showed deviations from the 2D-XY theory (especially at low temperatures), whereas the 3D-I model captured the data comprehensively. The Ising character of the magnetic correlations likely results from the orthorhombic anisotropy, which eventually drives the magnetic transition at \Sub T{N,1}. We remark, however, that in the $\Gamma$ \vs $\kappa$ plot, 3D-I scaling and the 2D-XY model were indistinguishable, due to identical dynamical critical exponents of the universality classes ($z$ = 2). 
Consequently, the 3D-I model provides the most conclusive description of the critical behaviors in \bicro, although a partial 2D-XY character can not be excluded. This ambiguity likely reflects the geometry of the exchange bonds in the bilayer structure of \bicro, which is intermediate between 2D and 3D. The theoretical description of the resulting crossover phenomena and detailed comparison with the experimental data are important challenges for future research.

In conclusion, our study of the critical magnetic correlations has confirmed \cro as a realization of the 2D-XY AFM on a square lattice. Along with Sr$_2$IrO$_4$, which hosts a nearly ideal 2D Heisenberg AFM \cite{Fujiyama.2012,Vale.2015}, this demonstration illustrates the power of 4$d$- and 5$d$-electron materials with strongly spin-orbit-entangled magnetic moments as a platform for fundamental research on quantum magnetism.

%-------------------------------------------
% Acknowledgements

\begin{acknowledgements}
Financial support by the European Research Council (Com4Com Advanced Grant No. 669550) and from the Deutsche Forschungsgemeinschaft (TRR80 Project No. 107745057) is gratefully acknowledged. Experiments were conducted at TRISP at FRM II in Garching, at FLEXX at BER II in Berlin, and at ThALES at ILL in Grenoble. We acknowledge helpful discussions with J. Bertinshaw, J. Porras, H.-A. Krug von Nidda, and P. Steffens.
\end{acknowledgements}

%-------------------------------------------
% APPENDIX

\appendix

\section{Energy integration for \bicro}
\label{app:energyIntegrationBicro}

\begin{figure}[htbp]
	\begin{center}
	\includegraphics[width=0.8\columnwidth]{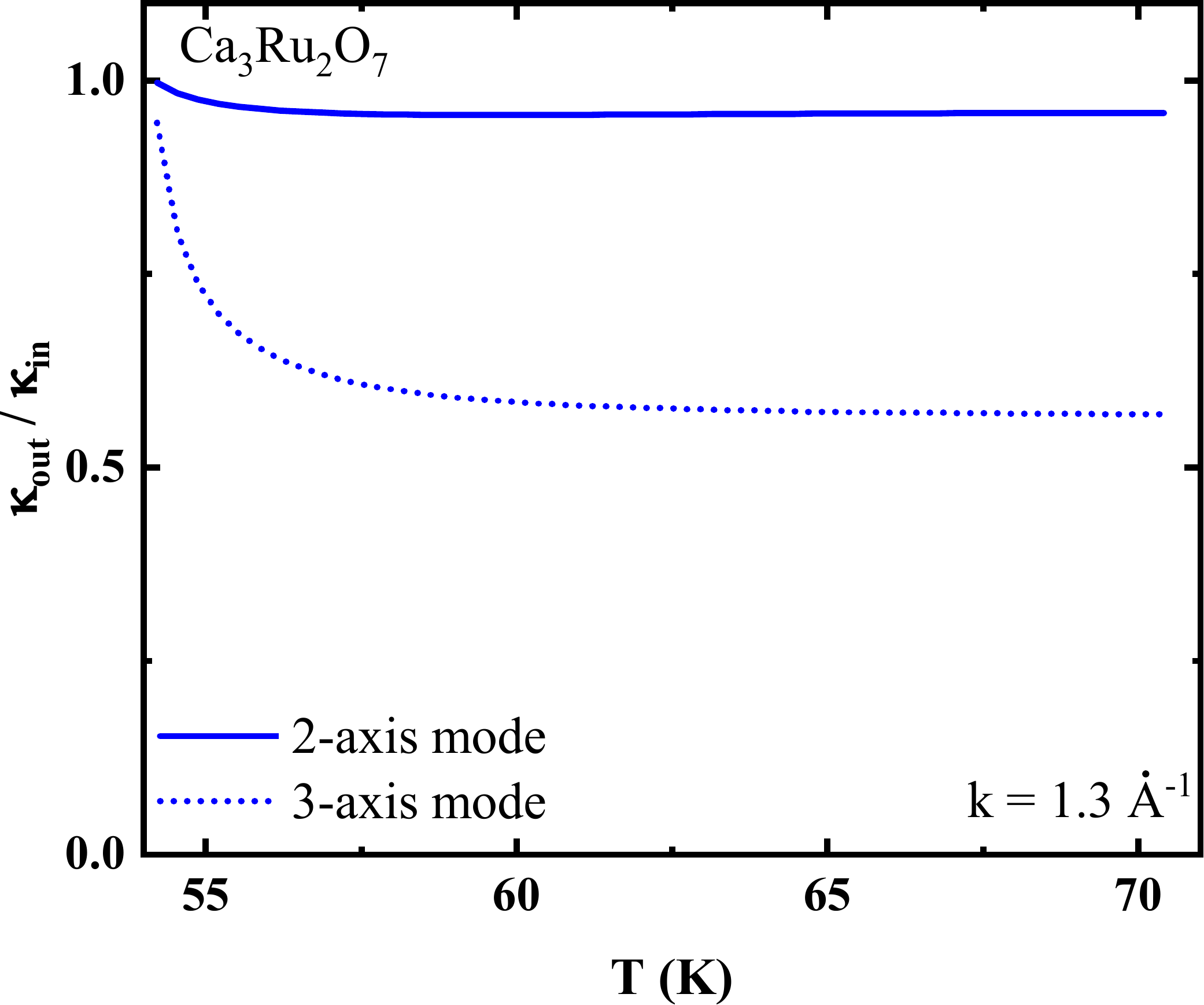}
		\caption{Simulated ratio of \Sub \kappa{out}/\Sub \kappa{in} vs. $T$ in the two- and three-axis mode, respectively (see details in text). }
		\label{fig:supp:TAS_E_integration}
	\end{center}
\end{figure}

Due to the 3D character of the AFM order in \bicro \cite{Bertinshaw.2021}, the ideal energy-integration configuration with \mbox{$k_f \parallel c$} in general cannot be obtained, as $\Sub QL$ cannot be chosen arbitrarily. To estimate the effects of inelasticity on the experimental $Q$-width we performed a numerical simulation, assuming an intrinsic $\Sub\kappa{in}$ and calculate $\Sub\kappa{out}$, which nominally corresponds to the width in the experimental $\Sub QH$-scans. For the simulation we assume a Lorentzian $S(q) = 1/[1 + (q/\Sub\kappa{in}(T))^2]$ with a $T$-dependent $Q$-width $\Sub\kappa{in}(T) = \kappa_0 t^{\nu}$, with $\nu = 0.5$ and $\kappa_0 = 0.2\,\AA^{-1}$. Note that this choice of parameters is close to experimental values extracted from PL-fitting in the main text. Further, we assume a Lorentzian $S(\omega) = \Gamma_q/[\Gamma_q^2 + \omega^2]$ with a $q$-dependent energy width $\Gamma_q = \Gamma(T) [1 + (q/\Sub\kappa{in}(T))^2]$ \cite{Schulhof.1970, Tucciarone.1971, Steffens.2011}. Here, the $T$-dependent energy-width is $\Gamma(T) = \Gamma_0 t^{z\nu}$, with $\Gamma_0 = 1$\,meV and $z\nu = 1$. We then calculate the integral [Eqn.~\eqref{equ:ScatTheo:approxBirg}] in the limits between $-\Sub kBT$ and $E_i$ for each $q_i$ in the \Sub QH-scan and fit the resulting intensity with a Lorentzian with HWHM $\Sub \kappa{out}$. The results of this simulation are shown in \fig{\ref{fig:supp:TAS_E_integration}}. Notably, for the two-axis mode, the reduction of the measured $\Sub\kappa{out}$ compared to the intrinsic $\Sub\kappa{in}$ is only of the order of a few percent. Such a change in the scaling behavior lies within our statistical error of the critical exponent $\nu$.  Hence, we did not correct the two-axis data of \bicro for the integration effect.

For comparison, we conducted the same simulation also for the triple-axis case [\fig{\ref{fig:supp:TAS_E_integration}}], where we added a Gaussian-distribution with the energy-resolution ($\pm \SI{0.04}{\milli\electronvolt}$ for $k_i = \kvec{1.3}$) as FWHM. As expected, for the latter case the difference between $\Sub\kappa{in}$ and $\Sub\kappa{out}$ is substantially larger.

\section{Critical scattering amplitude of \cro below \Sub TN}
\label{app:Criticalamplitudes}

Scaling theory predicts for the magnetic susceptibility PL behavior on both sides of \Sub TN for an ideal second-order phase transition. Specifically, $\X = A_-|t|^{-\gamma'}$ for $T < \Sub TN$ and $\X = A_+|t|^{-\gamma}$ with $\gamma = \gamma'$. 
The resulting critical exponent $\gamma = 0.426(7)$ is slightly reduced compared to the exponent extracted from $A_+|t|^\gamma$ in the range $110-140$\,K ($\gamma = 0.47(2)$), and still far from any universality class. This further supports the notion that the PL fit is not a proper choice for the description of the critical behavior in \cro.

\begin{figure}[htbp]
     \begin{center}
\includegraphics[width=0.8\columnwidth]{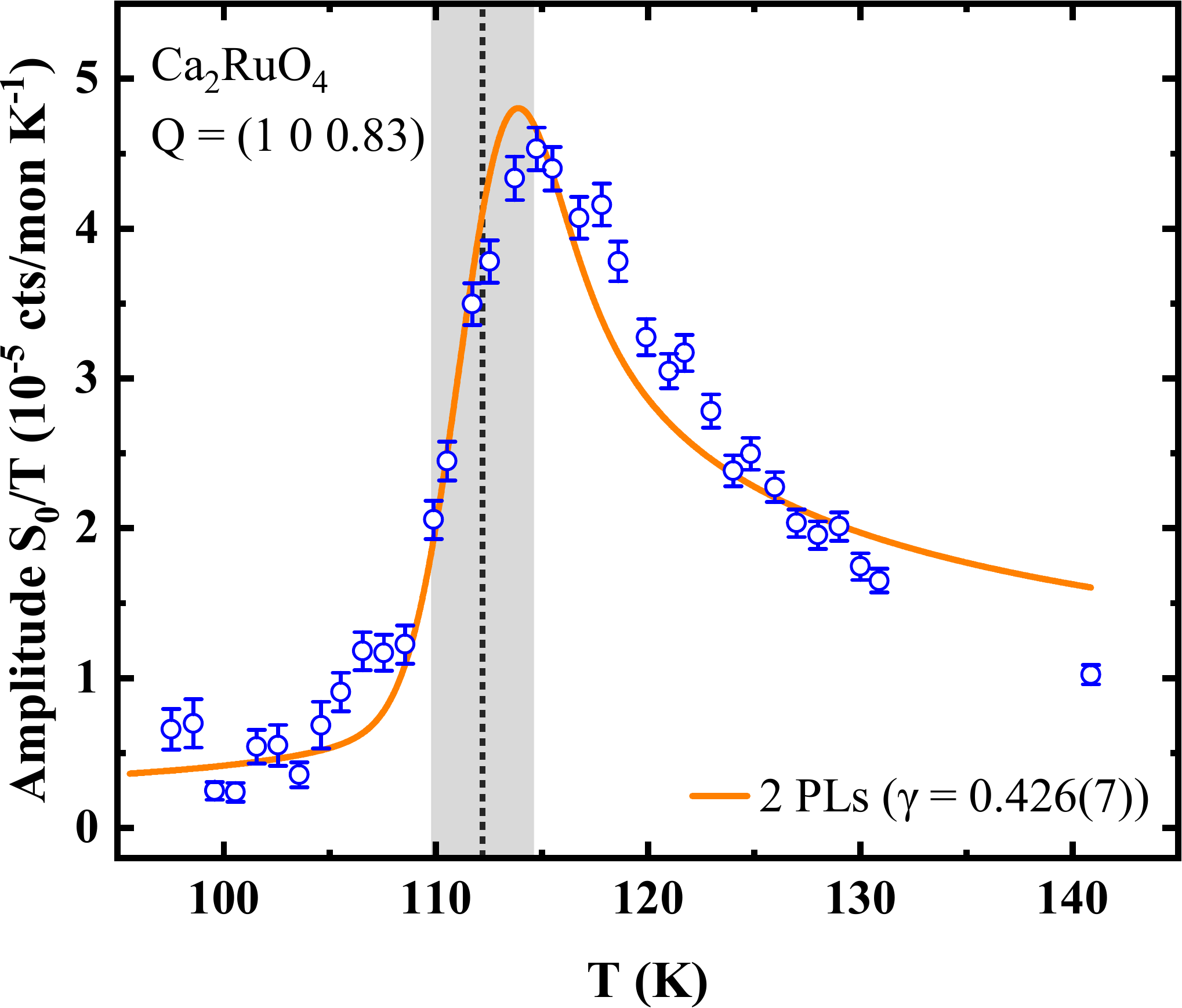}
         \caption{Peak amplitude $S_0(T)/T$ of \cro. The orange line corresponds to a PL fit, with $A_-|t|^\gamma$ and $A_+|t|^\gamma$ for $T < \Sub TN$ and $T > \Sub TN$, respectively. $A_-$ and $A_+$ are the universal amplitudes. The \Sub TN distribution was also taken into account. The extracted critical exponent $\gamma = 0.426(7)$ is close to the value obtained in the main text, where only $A_+|t|^\gamma$ was considered. The black vertical line indicates \Sub TN and the grey bar the variance of \Sub TN.}
         \label{fig:supp:CRO214_S0_2PLs}
     \end{center}
\end{figure}

\section{Sharp background peak of \bicro}
\label{app:sharpPeakBicro}

\begin{figure}[htbp]
	\begin{center}
	\includegraphics[width=0.8\columnwidth]{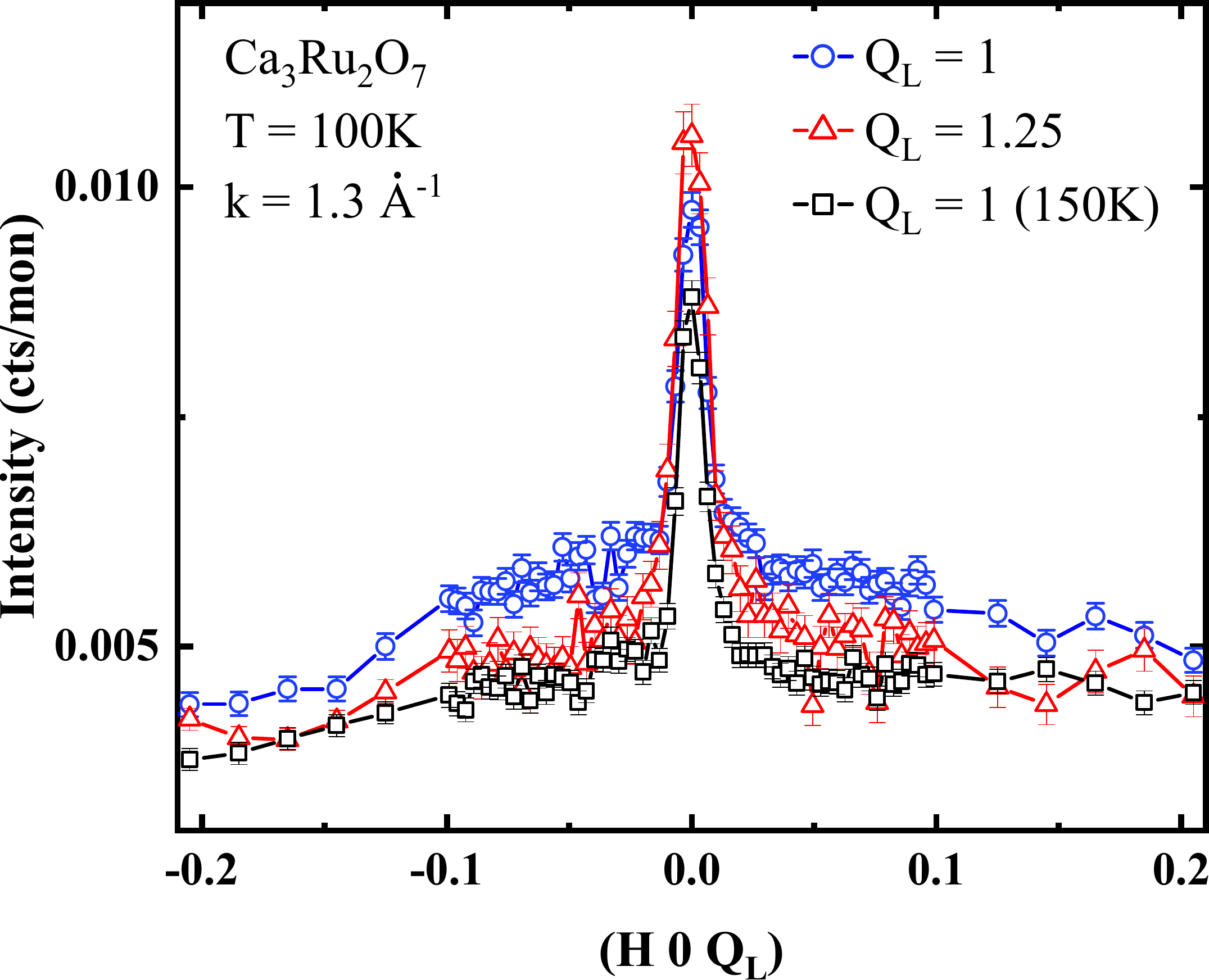}
		\caption{\peak{H}{0}{\Sub QL} scans in the two-axis mode at \Sub QL = 1 and 1.25 for $T$ = 100 and \SI{150}{\K}. The corresponding TAS angles are: A4 (A3) = 13.8\,° (83.1\,°) and 17.3\,° (81.3\,°).}
		\label{fig:supp:CRO327_QH_rod}
	\end{center}
\end{figure}

Figure.~\ref{fig:supp:CRO327_QH_rod} shows that a sharp peak at $H = 0$ is present in the scans around \peak{H}0{\Sub QL}. This sharp peak is independent of $T$ and \Sub QL, and thus we assign it to 2D diffuse nuclear scattering from disorder along the $c$-axis direction in \bicro.

\section{Energy-scan intensities of \cro}
\label{app:EscanIntensitiesCro}

\begin{figure}[htbp]
	\begin{center}
	\includegraphics[width=0.8\columnwidth]{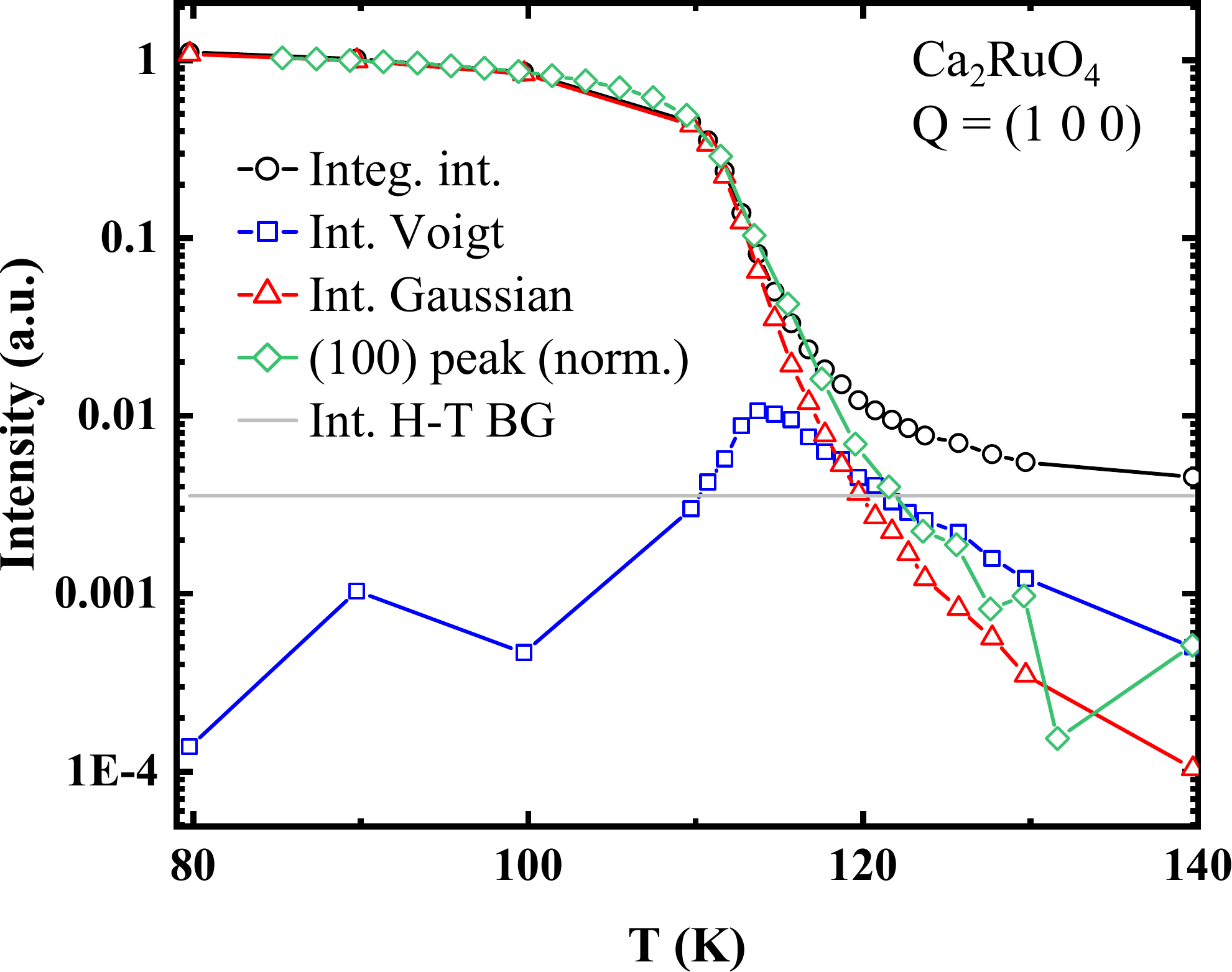}
		\caption{Integrated intensities of the fit components in the $E$-scans at $Q = \peak 100$ vs. $T$, including the overall integrated intensity (black), the Voigt-profile of the critical scattering (blue), and the resolution-limited Gaussian profile (red). For comparison, we also show the \peak100 peak intensity (green), measured with $k_f = 1.5\,\AA^{-1}$ and normalized on the \SI{80}{\K} point of the overall integrated intensity (black). The horizontal gray line indicates the intensity level of the high-temperature (H-T) BG at 170 K. }
		\label{fig:supp:CRO214_Flexx_Iint}
	\end{center}
\end{figure}

Figure~\ref{fig:supp:CRO214_Flexx_Iint} displays the integrated intensities of the fit components deduced from the energy-scans of \cro at FLEXX. Upon heating, the intense sharp resolution limited peak decreases significantly while above $\sim\SI{116}{\K}$ the broad critical component is gradually taking over almost all the spectral weight besides the elastic component. For comparison we added the \peak100 peak intensity measured with $k=$\kvec{1.5} and normalized on the \SI{80}{\K} point of the entire integrated intensity. The scaling of the normalized \peak100 peak resembles that of the sharp component, and thus we assign the latter to elastic magnetic scattering from the \peak100 still present above $\Sub TN$ due to the broad transition.

\section{Peak shape of $Q$-scans}
\label{app:PeakShape}

\begin{figure}[htbp]
	\begin{center}
	\includegraphics[width=0.8\columnwidth]{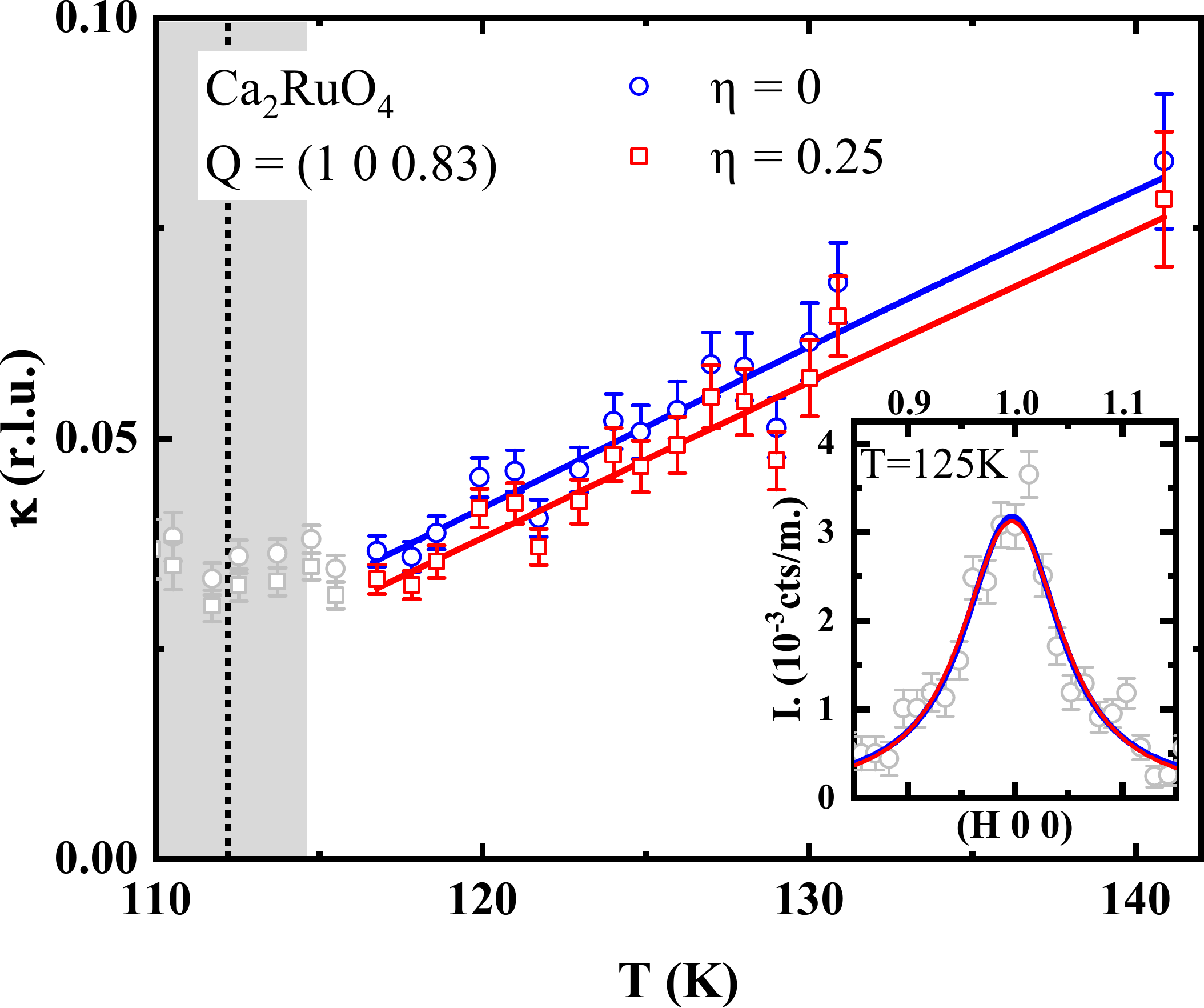}
		\caption{Comparison of the $Q$-widths $\kappa$ for a fitting of the $Q$-scans of \cro with $\eta = 0$ (blue points) and $\eta = 0.25$ (red points) [Eqn.~\eqref{equ:methods:Fisher}]. The grey points were not included in the fits. The black vertical line indicates \Sub TN and the grey bar the variance of \Sub TN. 
		The inset shows the corresponding fits at $T = 125$\,K.}
		\label{fig:supp:compare_eta}
	\end{center}
\end{figure}

As described in the Methods section, for the analysis of the $Q$-scans, we fitted the data with a Voigt-profile, \ie a convolution of a Gaussian (instrumental resolution) and a simple Lorentzian function [Eqn.~\eqref{equ:methods:Ornstein}]. Beyond this approximation, which is commonly used to describe critical scattering \cite{Klyushina.2021,Vale.2015,Vale.2019}, subtle deviations from the Lorentzian form were proposed, accounting for the critical exponent $\eta$ \cite{Fisher.1967, Collins.1989}: 
\begin{equation}
\X'(\Vb q)= \frac{\X'(\Vb 0)}{[1 + q^2 (1 - \eta/2)^{-1}/\kappa^2]^{1-\eta/2}}, 
\label{equ:methods:Fisher}
\end{equation}
with $\eta \approx 0$ and $\eta = 0.25$ predicted for 3D and 2D universality classes \cite{Collins.1989}, respectively. 
Hence, the peak-shape of the $Q$-scans also includes information on the critical behavior of the system. For bilayer \bicro, the 3D character of the critical fluctuations above \Sub TN suggests $\eta \approx 0$ consistent with the well-matching Voigt-fits [\fig{\ref{fig:results:QHscans_CRO327_examples}}]. For single-layer \cro, one would generally expect deviations from a simple Voigt-function due to the 2D character of the fluctuations.  However, we find that in our case fits with $\eta = 0$ and $\eta = 0.25$ yield peak shapes that are essentially the same within the experimental error of our data [insets in \fig{\ref{fig:supp:compare_eta}}]. Nevertheless, we carried out the analysis of the full set of $Q$-scans not only for $\eta = 0$ (Ornstein-Zernike form, Eqn.~\eqref{equ:methods:Fisher}), but also for $\eta = 0.25$ (2D-XY model, \cite{Kosterlitz.1974}). As can be seen in \fig{\ref{fig:supp:compare_eta}}, the effect of $\eta$ on our $Q$-widths is small, and is negligible with respect to the corresponding scaling behavior of $\kappa(T)$. Specifically, we obtain from 2D-XY fits [Eqn.~\eqref{equ:models:2DXY_kappa}] on the non-Lorentzian data ($\eta = 0.25$) for the Kosterlitz-Thouless temperature $\Sub T{KT} = 88(2)$\,K for \cro, which coincides within the statistical errors with the value discussed in the main text. 
Furthermore, we remark that in general a variety of other factors such as a non-Gaussian instrumental resolution and surface effects \cite{Papoular.1997,Cowley.2006} can also influence the peak-shape, which generally make the determination of $\eta$ challenging. 

%-------------------------------------------
% Bibliography

\bibliography{citavi_PHDthesis_corr}

\end{document}